\documentclass[aps,prd,onecolumn,nofootinbib]{revtex4-2}

\usepackage{amsmath,amssymb,amsfonts}
\usepackage{mathtools}
\usepackage{graphicx}
\usepackage{booktabs}
\usepackage{multirow}
\usepackage{bm}
\usepackage{xcolor}
\usepackage{hyperref}
\usepackage{placeins}
\hypersetup{colorlinks=true,linkcolor=blue,citecolor=blue,urlcolor=blue}

\makeatletter
\@addtoreset{equation}{section}
\renewcommand{\theequation}{\arabic{section}.\arabic{equation}}
\makeatother

\newcommand{\GeV}{\mathrm{GeV}}

\newcommand{\BR}{\mathcal B}

\begin{document}

\title{Hyperfine Mixing and Final-Spin Redistribution in Inclusive
$\Xi_{bc}\to H_{cc}+X$}
\author{Ishtiaq Ahmed}
\email{ishtiaq.ahmed@ncp.edu.pk}
\affiliation{National Center for Physics, Islamabad, Pakistan.}
\date{September 2, 2026}

\begin{abstract}
In this work, we study the effect of hyperfine mixing on the inclusive weak
transition $\Xi_{bc}\to H_{cc}+X$, which has been proposed as a useful channel
for searching for beauty-charmed baryons. Starting from the inclusive
heavy-diquark framework, we include mixing between the scalar and axial-vector
$bc$ configurations and decompose the final doubly charmed system into the
$J_f=1/2$ and $J_f=3/2$ spin sectors. Within a minimal scalar-to-axial
transition model, and for a benchmark scalar-channel normalization of order
unity at zero recoil, we find that the interference between the scalar and
axial components is large in the separate spin sectors but cancels after
summing over the two members of the adopted ground-state $ccq$ spin multiplet,
$\Gamma_{AS}^{1/2}=-\Gamma_{AS}^{3/2}$. Consequently, the spin-summed
inclusive rate is less sensitive to hyperfine mixing than the direct
ground-state contribution, provided feed-down from the spin-excited
$\Xi_{cc}^{*}$ partner is experimentally unresolved. For a scalar-channel
normalization of order unity, the direct contribution can be strongly spin selective:
$\Xi_{bc}^{(1)}$ dominantly feeds the $J_f=1/2$ $\Xi_{cc}$ channel, whereas
$\Xi_{bc}^{(2)}$ dominantly feeds the $J_f=3/2$ $\Xi_{cc}^{*}$ sector. We test
this pattern for several convention-matched literature admixtures and for a
separately labeled decay-interference diagnostic angle, together with the
dependence on the scalar-to-axial normalization and recoil-shape choice. The cancellation of the
spin-summed interference is a consequence of baryon spin recoupling within
the spectator and common-multiplet assumptions, whereas the magnitude of the
spin selectivity is conditional on the magnitude and relative phase of the
scalar-to-axial transition.
\end{abstract}
\maketitle

\section{Introduction}
\label{sec:introduction}

The spectroscopy and weak decays of doubly heavy baryons provide an important
laboratory to study the dynamics of QCD in systems containing more than one
heavy quark. Representative theoretical studies of doubly heavy baryon
spectroscopy, production, lifetimes, and weak transitions can be found in
Refs.~\cite{KiselevLikhoded:2002,EbertFaustovGalkin:2002,GershteinEtAl:2000,RobertsPervin:2008,KarlinerRosner:2014,BrownEtAl:2014,ShahEtAl:CPC2016,ShahEtAl:EPJA2016,WengEtAl:2018}. More recent QCD-sum-rule calculations have determined the masses and
residues of spin-$1/2$ and spin-$3/2$ doubly heavy baryons in their
ground and excited states
\cite{ShekariTousi:2024mso,ShekariTousi:2025xox}.
Related QCD-sum-rule studies of semileptonic transitions from doubly
charmed or doubly bottom baryons to singly heavy baryons can be found
in Refs.~\cite{Tousi:2024usi,ShekariTousi:2025fjf}. In such hadrons,
the two heavy quarks move relatively slowly with respect to each other and may
form a compact color-antitriplet diquark. This feature makes doubly heavy
baryons qualitatively different from singly heavy baryons and heavy mesons,
and allows one to test ideas based on heavy-quark spin symmetry,
nonrelativistic QCD, and heavy-diquark effective theory
\cite{GeorgiWise:1990,SavageWise:1990,WhiteSavage:1991,HuMehen:2006,MehenPowell:2011,IsgurWise:1989,IsgurWise:1991,EichtenHill:1990,Grinstein:1990,Luke:1990,ManoharWise:1994,Neubert:1994,ManoharWise:2000,CaswellLepage:1986,BodwinBraatenLepage:1995,BrambillaEtAl:2005,BrambillaEtAl:2011}.

The experimental observation of doubly charmed baryons has opened a new
window on the study of heavy-hadron spectroscopy. The observation of the
$\Xi_{cc}^{++}$ baryon by the LHCb Collaboration established the first
unambiguous doubly charmed baryon signal~\cite{LHCb:2017XiccObservation}, following earlier
SELEX evidence~\cite{SELEX:2002Xicc,SELEX:2005Xicc}. The state was subsequently observed in
additional decay modes~\cite{LHCb:2018XiccDecay}. Its lifetime has also been measured by
LHCb, and searches for the isospin partner $\Xi_{cc}^{+}$ have been performed
\cite{LHCb:2018XiccLifetime,LHCb:2020XiccPlusSearch}. More recently, LHCb has reported the observation of
$\Xi_{cc}^{+}$ with the Run-3 detector~\cite{LHCb:2026XiccPlusObservation}. The experimental program
on doubly charmed baryons therefore continues to motivate the study of doubly
heavy systems containing charm and bottom quarks. In particular, the
beauty-charmed baryons $\Xi_{bc}$ and $\Omega_{bc}$ remain especially
interesting because they contain two different heavy flavors.

Among the proposed discovery modes, the inclusive transition
\begin{equation*}
\Xi_{bc}\to\Xi_{cc}^{++}+X
\end{equation*}
is particularly attractive. The produced $\Xi_{cc}^{++}$ is itself weakly
decaying and can provide a displaced secondary vertex, reducing combinatorial
background. In the inclusive approach of Qin \emph{et al.},
$\Xi_{bc}^{+}\to\Xi_{cc}^{++}+X$ was studied in a heavy-diquark framework and
was proposed as a promising discovery channel~\cite{QinEtAl:2022}. Weak decays and
lifetimes of doubly heavy baryons have also been studied in a variety of
quark-model, effective-theory, and inclusive approaches
\cite{KiselevEtAl:1999,KiselevEtAl:2000,FaesslerEtAl:2001,HernandezEtAl:2008,FaesslerEtAl:2009,AlbertusEtAl:2007,BigiUraltsevVainshtein:1992,BigiEtAl:1993,NeubertSachrajda:1997,GuberinaMelicStefancic:1999,ChengShi:2018}.
Weak transitions into spin-$1/2$ final baryons have also been analyzed in the
light-front approach~\cite{WangYuZhao:2017}, while strong and radiative decays
of ground and excited doubly charmed baryons have been investigated in
Refs.~\cite{XiaoEtAl:2017,XiaoLuZhu:2018}.
Experimentally, direct searches for beauty-charmed baryons remain challenging;
for example, LHCb has searched for $\Xi_{bc}^{+}\to J/\psi\,\Xi_c^+$ and set
upper limits in the absence of a definitive signal~\cite{LHCb:2023XibcSearch}. Related
searches and production estimates are discussed in Refs.~\cite{BerezhnoyEtAl:1998,Baranov:1996,BerezhnoyEtAl:2018,ChangEtAl:2006,JiangQiao:2012,ZhangEtAl:2011}.

The existing inclusive estimate treats the dominant contribution as an
axial-vector $bc$ diquark changing into an axial-vector $cc$ diquark. Physical
$bcq$ baryons, however, are not necessarily pure scalar or pure axial-vector
heavy-diquark states. Since the heavy quarks have different flavors, the
$S_{bc}=0$ and $S_{bc}=1$ configurations can mix through hyperfine
interactions. Such mixing has been studied in quark models and heavy-quark
spin-symmetry analyses and can significantly modify spin-resolved decay rates
\cite{RobertsPervin:2008,RobertsPervin:2009,AlbertusEtAl:2010a,AlbertusEtAl:2010b,AlievAziziSavci:2012,EbertFaustovGalkin:2004}. We write
\begin{equation*}
\begin{aligned}
|\Xi_{bc}^{(1)}\rangle&=\cos\theta\,|A\rangle-\sin\theta\,|S\rangle,\\
|\Xi_{bc}^{(2)}\rangle&=\sin\theta\,|A\rangle+\cos\theta\,|S\rangle,
\end{aligned}
\end{equation*}
where $|A\rangle\equiv|S_{bc}=1\rangle$ and
$|S\rangle\equiv|S_{bc}=0\rangle$.

In this work, we revisit $\Xi_{bc}\to H_{cc}+X$ by including hyperfine mixing
in the initial baryon. Throughout this work, $H_{cc}$ denotes the doubly
charmed hadronic system generated by the $cc$ diquark. The spin-resolved
analysis is restricted to the spectator-preserving $S$-wave $ccq$ channel in
which the light cloud is represented by a light quark with $s_q=1/2$, giving
the two ground-state spin partners $J_f=1/2$ and $J_f=3/2$. The new element is
not hyperfine mixing itself, but its incorporation into the inclusive
heavy-diquark discovery channel together with explicit recoupling of the
heavy-diquark and light-quark spins and a decomposition into the two final-spin
sectors,
\begin{equation*}
S_{cc}=1,\qquad s_q=\frac12,\qquad
1\otimes\frac12=\frac12\oplus\frac32.
\end{equation*}
Thus the final system contains a ground-state $\Xi_{cc}(J_f=1/2)$ component
and a spin-excited $\Xi_{cc}^{*}(J_f=3/2)$ component.


It is important to mention here that hyperfine mixing in doubly heavy $bcq$
baryons and its effects on exclusive weak-decay widths have been studied
previously. Roberts and Pervin found that mixing can considerably modify the
relative decay rates into the $\Xi_{cc}$ and $\Xi_{cc}^{*}$ final states and
can also affect their summed semileptonic rate~\cite{RobertsPervin:2009}. Albertus,
Hern\'andez and Nieves subsequently confirmed the strong sensitivity of the
spin-resolved $b\to c$ semileptonic widths to the physical admixtures
\cite{AlbertusEtAl:2010a,AlbertusEtAl:2010b}. They also discussed how heavy-quark spin symmetry can be
used to obtain information about the mixing coefficients. Therefore, neither
hyperfine mixing itself nor the general redistribution of decay strength
between the $J_f=1/2$ and $J_f=3/2$ channels is regarded as a new result of the
present work.

Recently, Endo \emph{et al.} formulated heavy-quark sum rules by using a
spin-decomposition framework~\cite{EndoEtAl:2025}. In their analysis, the sum over the
allowed final-spin partners makes it possible to use the orthogonality of the
Clebsch--Gordan coefficients and obtain relations among the squared decay
amplitudes. In this context, it is useful to note that the general
spin-completeness mechanism associated with the final-spin sum is not
particular to the decay considered here.

In the present work, our purpose is to study this problem in the inclusive
heavy-diquark discovery framework of Qin \emph{et al.} For this purpose, we
extend the original $A_{bc}\to A_{cc}$ calculation by introducing the
scalar-to-axial transition $S_{bc}\to A_{cc}$. The two amplitudes are combined
coherently for the physical hyperfine-mixed $\Xi_{bc}$ states and are then
recoupled into the physical $J_f=1/2$ and $J_f=3/2$ baryon sectors. This
construction includes both the semileptonic and nonleptonic partonic $X$
channels entering the inclusive rate.

Within the spectator and common-multiplet assumptions, we find
\begin{equation}
\Gamma_{AS}^{1/2}=-\Gamma_{AS}^{3/2}.
\label{eq:comparison-interference}
\end{equation}
One can see from Eq.~\eqref{eq:comparison-interference} that the
scalar--axial interference is nonzero in the separate final-spin sectors but
cancels after the two members of the adopted ground-state $ccq$ multiplet are
summed. This relation concerns the interference contribution only. The
diagonal $AA$ and $SS$ terms remain in the complete rate, and corrections to
the aforementioned assumptions can generate a residual spin-summed
interference.

The physical interpretation of this result depends on whether the two
final-spin sectors can be separated experimentally. If the $\Xi_{cc}^{*}$
contribution feeds down to $\Xi_{cc}$ without being resolved, the observed
sample approximately combines the direct $J_f=1/2$ contribution and the
feed-down $J_f=3/2$ contribution. Consequently, the spin-summed yield can be
less sensitive to hyperfine mixing even when the separate final-spin channels
show a sizeable redistribution of decay strength. The main differences
between the present analysis and the aforementioned studies are summarized in
Table~\ref{tab:previous-work}.

\begin{table}[t]
\caption{\label{tab:previous-work}
Comparison of the present analysis with previous hyperfine-mixing and
heavy-quark spin-sum studies.}
\begin{ruledtabular}
\begin{tabular}{p{0.14\textwidth}p{0.21\textwidth}
                p{0.23\textwidth}p{0.24\textwidth}}
Study
&
Decay framework
&
Main result
&
Relation to the present analysis
\\[2pt]
\hline

Roberts and Pervin~\cite{RobertsPervin:2009}
&
Exclusive semileptonic
$\Xi_{bc}\to\Xi_{cc}^{(*)}\ell\bar{\nu}$
&
Hyperfine mixing considerably changes the resolved decay widths and can also
affect their sum
&
Establishes the importance of spin redistribution, but does not consider the
Qin inclusive discovery framework
\\[3pt]

Albertus, Hern\'andez and Nieves~\cite{AlbertusEtAl:2010a,AlbertusEtAl:2010b}
&
Exclusive $b\to c$ semileptonic decays and heavy-quark spin symmetry
&
The spin-resolved widths are sensitive to the physical hyperfine admixtures
&
Provides a literature-motivated mixing example, but does not analyze the
inclusive scalar--axial interference considered here
\\[3pt]

Endo \emph{et al.}~\cite{EndoEtAl:2025}
&
Heavy-quark sum rules based on spin decomposition
&
The sum over final-spin partners allows the use of Clebsch--Gordan
orthogonality
&
Provides the related general spin-completeness mechanism, but does not apply
it to hyperfine-mixed inclusive $\Xi_{bc}$ decay
\\[3pt]

Present work
&
Inclusive $\Xi_{bc}\to H_{cc}+X$, including semileptonic and nonleptonic
partonic channels
&
The scalar--axial interference satisfies
$\Gamma_{AS}^{1/2}=-\Gamma_{AS}^{3/2}$
&
Connects the spin-resolved cancellation with the distinction between direct
production and the yield including unresolved feed-down
\end{tabular}
\end{ruledtabular}
\end{table}

We also test the dependence on the hyperfine mixing angle, scalar-to-axial
transition normalization, and recoil shape. In particular, we compare the
normalized recoil dependence obtained from Qin's form factors with the
explicit wave-function-overlap treatment employed in the nonrelativistic
potential-model approach of Ridgway and Wise and subsequently applied by Yang
to $\Xi_{bcq}\to\Xi_{ccq}+X$~\cite{RidgwayWise:2019,Yang:2026}. The normalized-shape
replacement changes the integrated rate only mildly when the Qin
normalization is kept fixed; it is a shape-sensitivity test rather than a
comparison of absolute model normalizations.
Representative heavy-quark potential constructions and reviews underlying
this class of nonrelativistic calculations can be found in
Refs.~\cite{EichtenEtAl:1975,EichtenEtAl:1978,QuiggRosner:1979,BuchmullerTye:1981,Martin:1980,GodfreyIsgur:1985,LuchaSchoberlGromes:1991}.

\section{Inclusive framework for \texorpdfstring{$\Xi_{bc}\to H_{cc}+X$}{Xi_bc -> H_cc + X}}
\label{sec:formalism}

We follow the heavy-diquark picture in which the two heavy quarks form a
compact color-antitriplet subsystem. In the leading approximation, the light
quark acts as a spectator and the weak transition is described by the decay
of the heavy $bc$ diquark into a $cc$ diquark, as in Ref.~\cite{QinEtAl:2022}.

With $P_L=(1-\gamma_5)/2$, the semileptonic effective Hamiltonian is
\begin{equation}
\mathcal H_{\rm eff}^{\rm sl}=\frac{4G_F}{\sqrt2}V_{cb}
(\bar c\gamma_\mu P_L b)(\bar\ell\gamma^\mu P_L\nu_\ell)+\mathrm{h.c.},
\label{eq:heff-sl}
\end{equation}
while for nonleptonic modes
\begin{equation}
\mathcal H_{\rm eff}^{\rm nl}=\frac{4G_F}{\sqrt2}V_{cb}V_{UD}^{*}
\left[C_1(\mu)O_1^{UD}+C_2(\mu)O_2^{UD}\right]+\mathrm{h.c.},
\label{eq:heff-nl}
\end{equation}
with
\begin{align*}
O_1^{UD}&=(\bar c_i\gamma_\mu P_L b_i)(\bar D_j\gamma^\mu P_L U_j),\\
O_2^{UD}&=(\bar c_i\gamma_\mu P_L b_j)(\bar D_j\gamma^\mu P_L U_i).
\end{align*}
Here $i,j$ are color indices. The factor four is compensated by the two
projectors $P_L$ and makes the factor $1/2$ in the left-handed hadronic current
explicit.

We use the same heavy-diquark mass approximation as the Qin baseline,
\begin{equation*}
M_i=m_b+m_c,\qquad M_f=2m_c.
\end{equation*}
Accordingly, the recoil variable, $q^2$ endpoint, and phase-space factors are
evaluated at the heavy-diquark level. Binding-energy differences and
hyperfine mass splittings of the physical baryons are not included in this
baseline kinematics. With $q=P_i-P_f$,
\begin{equation*}
w=\frac{M_i^2+M_f^2-q^2}{2M_iM_f},\qquad
k(q^2)=\frac{\sqrt{\lambda(M_i^2,M_f^2,q^2)}}{2M_i},
\end{equation*}
where
\begin{equation*}
\lambda(a,b,c)=a^2+b^2+c^2-2ab-2ac-2bc.
\end{equation*}
The physical range is $m_\ell^2\le q^2\le q^2_{\rm max}$ with
$q^2_{\rm max}=(M_i-M_f)^2$.

For the axial-vector $bc\to cc$ diquark transition, following Ref.~\cite{QinEtAl:2022},
\begin{align}
V_A^\mu&=\sqrt{2M_iM_f}\left[-a_0(\epsilon_f^*\!\cdot\epsilon_i)v_i^\mu
-a_1(\epsilon_f^*\!\cdot\epsilon_i)v_f^\mu
+a_2(\epsilon_f^*\!\cdot v_i)\epsilon_i^\mu
+a_3(v_f\!\cdot\epsilon_i)\epsilon_f^{*\mu}\right],\nonumber\\
A_A^\mu&=\sqrt{2M_iM_f}\left[-ib_0\epsilon^{\mu\nu\alpha\beta}
\epsilon_{f\nu}^*\epsilon_{i\alpha}v_{i\beta}
-ib_1\epsilon^{\mu\nu\alpha\beta}
\epsilon_{f\nu}^*\epsilon_{i\alpha}v_{f\beta}\right],
\label{eq:AA-current}
\end{align}
with $b_0=a_0$, $b_1=a_1$, and $a_3=a_2$. We define reduced currents by
stripping the common $\sqrt{2M_iM_f}$ factor. The left-handed reduced current
is
\begin{equation*}
\widehat H_A^\mu=\frac12(\widehat V_A^\mu-\widehat A_A^\mu),
\end{equation*}
so that
\begin{equation*}
\langle A_{cc}|\bar c\gamma^\mu P_L b|A_{bc}\rangle
=\sqrt{2M_iM_f}\,\widehat H_A^\mu.
\end{equation*}

The form factors are parametrized as
\begin{equation}
f(q^2)=\frac{f(0)}{1-q^2/m_{B_c}^2}
\left[1+b\,\zeta(q^2)+c\,\zeta^2(q^2)\right],
\label{eq:ff-param}
\end{equation}
with $\zeta(q^2)=z(q^2)-z(0)$,
\begin{equation*}
z(q^2)=\frac{\sqrt{t_+-q^2}-\sqrt{t_+-t_0}}
{\sqrt{t_+-q^2}+\sqrt{t_+-t_0}},\qquad
t_\pm=(M_i\pm M_f)^2,
\end{equation*}
and
\begin{equation*}
t_0=t_+\left(1-\sqrt{1-\frac{t_-}{t_+}}\right).
\end{equation*}

\begin{table}[t]
\centering
\caption{Input parameters used in the numerical analysis. The heavy-diquark
masses follow the Qin baseline. Mixing scenarios are listed separately in
Table~\ref{tab:mixing-scenarios}.}
\label{tab:inputs}
\begin{tabular}{lclc}
\toprule
Parameter&Value&Parameter&Value\\
\midrule
$G_F$&$1.1663787\times10^{-5}\,\GeV^{-2}$&$|V_{cb}|$&0.041\\
$V_{ud}$&0.974&$V_{us}$&0.225\\
$V_{cd}$&0.225&$V_{cs}$&0.973\\
$m_b$&4.749 GeV&$m_c$&1.392 GeV\\
$M_i$&6.141 GeV&$M_f$&2.784 GeV\\
$m_{B_c}$&6.2749 GeV&$q^2_{\rm max}$&11.2694 GeV$^2$\\
$m_e$&0&$m_\mu$&0.105658 GeV\\
$m_\tau$&1.77686 GeV&$C_1(m_b)$&1.10\\
$C_2(m_b)$&$-0.24$&$3C_1^2+2C_1C_2+C_2^2$&3.1596\\
$\tau(\Xi_{bc}^+)$&508 fs&$\hbar$&$6.582119569\times10^{-25}$ GeV s\\
\bottomrule
\end{tabular}
\end{table}

\begin{table}[t]
\centering
\caption{Form-factor parameters for the axial-vector $bc\to cc$ transition.}
\label{tab:ff-inputs}
\begin{tabular}{c|ccc}
\toprule
Form factor&$f(0)$&$b$&$c$\\
\midrule
$a_0$&0.124&$-52.9$&1898.2\\
$a_1$&0.191&$-57.0$&388.0\\
$a_2$&0.247&$-58.6$&$-238.2$\\
\bottomrule
\end{tabular}
\end{table}

With these inputs, the endpoint-constrained form factors $a_0$ and $a_1$
satisfy the expected zero-recoil normalization to very good accuracy. For
$a_2$, the published parameter table of Ref.~\cite{QinEtAl:2022} lists
$c_{a_2}=+238.2$. Direct insertion of that value into Eq.~\eqref{eq:ff-param}
with the masses used here gives $a_2(q^2_{\rm max})\simeq1.239$, whereas
$c_{a_2}=-238.2$ gives $a_2(q^2_{\rm max})\simeq0.9994$, consistent with the
endpoint condition used in our numerical reproduction. We therefore adopt
$c_{a_2}=-238.2$ as an explicit reproducibility choice rather than as a refit.
A controlled calculation with both signs is presented in
Sec.~\ref{sec:a2-sign-test}.

The spin-summed tensor for the unmixed axial channel is
\begin{equation}
T_A(q^2,m_\ell)=\frac{2M_iM_f}{3}
\sum_{\lambda_i,\lambda_f}\widehat H_A^\mu K_{\mu\nu}^{(\ell)}
\widehat H_A^{\nu *},
\label{eq:TA-tensor}
\end{equation}
where
\begin{equation}
K_{\mu\nu}^{(\ell)}=(1-r_\ell)^2\left[\frac{2+r_\ell}{2}
(q_\mu q_\nu-q^2g_{\mu\nu})+\frac{3r_\ell}{2}q_\mu q_\nu\right],
\qquad r_\ell=\frac{m_\ell^2}{q^2}.
\label{eq:leptonic-tensor}
\end{equation}
For $m_\ell\to0$, $K_{\mu\nu}=q_\mu q_\nu-q^2g_{\mu\nu}$. The differential
semileptonic width is
\begin{equation}
\frac{d\Gamma_\ell}{dq^2}=\frac{G_F^2|V_{cb}|^2}{48\pi^3M_i^3}
\sqrt{\lambda(M_i^2,M_f^2,q^2)}\,T_A(q^2,m_\ell).
\label{eq:sl-width}
\end{equation}

The hadronic modes are included through the partonic replacement adopted in
Ref.~\cite{QinEtAl:2022},
\begin{equation}
|V_{cb}|^2\longrightarrow|V_{cb}V_{UD}^{*}|^2
\left(3C_1^2+2C_1C_2+C_2^2\right),
\label{eq:hadronic-replacement}
\end{equation}
for $\bar ud$, $\bar us$, $\bar cd$, and $\bar cs$. For charm-containing
virtual-$W$ channels we use the massive-current template with $m_\ell\to m_c$.
In this prescription the other quark produced by the virtual $W$ is treated as
massless in the phase-space template. This is an inclusive partonic
approximation inherited from the Qin baseline and not a calculation of the
full nonleptonic spectral function, higher-order QCD corrections, or
additional nonperturbative power corrections.

\section{Hyperfine mixing and scalar-to-axial transition}
\label{sec:mixing}

Throughout this section, ``scalar'' and ``axial-vector'' label the
heavy-diquark spin configurations $S_{bc}=0$ and $S_{bc}=1$, respectively;
no Lorentz-scalar weak operator is introduced. The weak interaction remains
the Standard-Model left-handed charged current.

\subsection{Ordered spin basis and phase convention}

To lock the phase convention, we order the heavy-quark spin labels with the
active $b$ quark first and the spectator $c$ quark second. We define
\begin{align}
|S_{bc},0\rangle&=\frac{1}{\sqrt2}
\left(|\uparrow_b\downarrow_c\rangle-|\downarrow_b\uparrow_c\rangle\right),
\nonumber\\
|A_{bc},+1\rangle&=|\uparrow_b\uparrow_c\rangle,\qquad
|A_{bc},-1\rangle=|\downarrow_b\downarrow_c\rangle,\nonumber\\
|A_{bc},0\rangle&=\frac{1}{\sqrt2}
\left(|\uparrow_b\downarrow_c\rangle+|\downarrow_b\uparrow_c\rangle\right).
\label{eq:ordered-bc-spin-basis}
\end{align}
The active daughter charm quark is likewise written before the spectator
charm quark in the final $cc$ state. With the standard Condon--Shortley
Clebsch--Gordan phases, the nonrelativistic spatial axial current then gives
\begin{equation}
\sigma_b^z|S_{bc},0\rangle=+|A_{cc},0\rangle.
\label{eq:scalar-spin-sign}
\end{equation}
The plus sign in Eq.~\eqref{eq:scalar-spin-sign} fixes the spin-algebra sign
of the reference scalar transition amplitude $h_S^\mu$ in this ordered
basis. Reversing the order of the two heavy-spin labels changes the phase of
the antisymmetric singlet and must therefore be accompanied by the
corresponding change in the mixing coefficients and transition amplitude.

\subsection{Basis Conventions and Mixing Parameters}

Albertus, Hern\'andez and Nieves (AHN) quote the admixtures
\begin{equation*}
|\Xi_{bc}^{(1)}\rangle_{\rm AHN}=0.902|\Xi_{bc}'\rangle+0.431|\Xi_{bc}\rangle,
\qquad
|\Xi_{bc}^{(2)}\rangle_{\rm AHN}=-0.431|\Xi_{bc}'\rangle+0.902|\Xi_{bc}\rangle,
\end{equation*}
where $|\Xi_{bc}'\rangle$ and $|\Xi_{bc}\rangle$ denote their scalar and axial
basis states, respectively. The precise mapping to our convention is
\begin{equation*}
|A\rangle=|\Xi_{bc}\rangle_{\rm AHN},\qquad
|S\rangle=-|\Xi_{bc}'\rangle_{\rm AHN}.
\end{equation*}
Therefore
\begin{align*}
|\Xi_{bc}^{(1)}\rangle
&=0.431|A\rangle-0.902|S\rangle
=0.431|\Xi_{bc}\rangle_{\rm AHN}
+0.902|\Xi_{bc}'\rangle_{\rm AHN},\\
|\Xi_{bc}^{(2)}\rangle
&=0.902|A\rangle+0.431|S\rangle
=0.902|\Xi_{bc}\rangle_{\rm AHN}
-0.431|\Xi_{bc}'\rangle_{\rm AHN}.
\end{align*}
This reproduces both AHN states exactly, with
\begin{equation*}
\cos\theta_{\rm AHN}=0.431,\qquad
\sin\theta_{\rm AHN}=0.902,\qquad
\tan\theta_{\rm AHN}=\frac{0.902}{0.431},\qquad
\theta_{\rm AHN}\simeq64.46^\circ.
\end{equation*}
The angle is model dependent but provides a useful comparison point. The
previously implicit scalar-basis phase is thus fixed explicitly by
$|S\rangle=-|\Xi_{bc}'\rangle_{\rm AHN}$. The associated AHN masses are
$6.967$ GeV for $\Xi_{bc}^{(1)}$ and $6.919$ GeV for
$\Xi_{bc}^{(2)}$~\cite{AlbertusEtAl:2010b}; thus their state numbered ``(1)'' is the
heavier one.

For comparison, Roberts and Pervin quote the rounded spin wave functions
\cite{RobertsPervin:2009}
\begin{equation*}
|\Xi_{bc}\rangle_{\rm RP}=0.92|\chi^\lambda\rangle
+0.39|\chi^\rho\rangle,
\qquad
|\Xi_{bc}'\rangle_{\rm RP}=0.92|\chi^\rho\rangle
-0.39|\chi^\lambda\rangle,
\end{equation*}
where $\chi^\lambda$ and $\chi^\rho$ carry heavy-diquark spin one and zero,
respectively. Taking
\begin{equation*}
|A\rangle=|\chi^\lambda\rangle_{\rm RP},\qquad
|S\rangle=-|\chi^\rho\rangle_{\rm RP},
\end{equation*}
their first state agrees with our $|\Xi_{bc}^{(1)}\rangle$, while their second
state agrees with $-|\Xi_{bc}^{(2)}\rangle$. The corresponding angle is
\begin{equation*}
\theta_{\rm RP}=\tan^{-1}\!\left(\frac{0.39}{0.92}\right)
\simeq22.97^\circ,
\end{equation*}
up to the rounding of the published coefficients. The masses quoted with
these wave functions are $7.011$ and $7.047$ GeV, respectively, so the first
RP state is the lighter one.

Aliev, Azizi and Savc\i{} obtain the QCD-sum-rule angle
\cite{AlievAziziSavci:2012}
\begin{equation*}
\phi_{\Xi_{bc}}=16^\circ\pm5^\circ
\end{equation*}
for a conventional rotation ordered as
$(|\Xi_{bc}\rangle,|\Xi_{bc}'\rangle)$, with the unprimed and primed basis
states corresponding to the axial and scalar heavy-diquark configurations.
With $|A\rangle=|\Xi_{bc}\rangle$ and
$|S\rangle=-|\Xi_{bc}'\rangle$, this maps to
$\theta_{\rm QCDSR}=16^\circ\pm5^\circ$ in our convention, up to an irrelevant
overall phase of the second physical state. That analysis determines the
mixing-angle magnitude but does not assign the two physical masses needed to
order the decay states.

It is important to mention here that the nonrelativistic value
$\phi_{\Xi_{bc}}\simeq25.5^\circ$ quoted in Ref.~\cite{AlievAziziSavci:2012} uses the
complementary scalar-first convention. It corresponds to
$90^\circ-25.5^\circ\simeq64.5^\circ$ in our axial-first convention and is
numerically the same AHN admixture already represented by
$\theta_{\rm AHN}=64.46^\circ$. We therefore do not count these two angle
labels as independent scenarios.

It is useful to verify that this basis phase cannot affect a physical width.
For real mixing coefficients, write
\begin{equation*}
\mathcal M_i=a_i\mathcal M_A+b_i\mathcal M_S,
\qquad
\Gamma_i=a_i^2\Gamma_{AA}+b_i^2\Gamma_{SS}
+2a_ib_i\Gamma_{AS}.
\end{equation*}
Under the harmless basis rephasing $|S\rangle\to-|S\rangle$, the same
physical state and matrix element require
\begin{equation*}
b_i\to-b_i,\qquad \mathcal M_S\to-\mathcal M_S,
\qquad \Gamma_{AS}\to-\Gamma_{AS}.
\end{equation*}
The interference product is invariant,
\begin{equation}
2a_i(-b_i)(-\Gamma_{AS})=2a_ib_i\Gamma_{AS},
\label{eq:rephasing-invariance}
\end{equation}
and hence every reported width is unchanged. Changing only the sign of
$\mathcal M_S$ while keeping $b_i$ fixed is instead a different physical
assumption for the relative scalar--axial phase.

The value $\theta_{\rm B}=41.06^\circ$ is instead a decay-level diagnostic.
Before full baryon-spin recoupling, an auxiliary overlap for the second mixed
state has the schematic form
\begin{equation*}
I_{\Xi_2}^{\rm eff}(p,\theta)=I_1(p/2)\sin\theta-
\frac{\sqrt3}{2}I_0(p/2)\cos\theta.
\end{equation*}
If $I_0=I_1$, destructive interference occurs at
\begin{equation*}
\tan\theta=\frac{\sqrt3}{2},\qquad\theta=40.89^\circ.
\end{equation*}
Using explicit spin-dependent overlaps shifts the numerical minimum to
$41.05685^\circ$. We use $\theta_{\rm B}=41.06^\circ$ only as a diagnostic
point, not as a spectroscopy determination of the physical hyperfine
mixing angle.

The most general scalar-to-axial diquark matrix element can be written as
\begin{equation}
V_S^\mu=i\sqrt{2M_iM_f}\,g_V(q^2)
\epsilon^{\mu\nu\alpha\beta}\epsilon_\nu^*v_{f\alpha}v_{i\beta},
\label{eq:scalar-vector-current}
\end{equation}
and
\begin{equation}
A_S^\mu=\sqrt{2M_iM_f}\left[f(q^2)\epsilon^{*\mu}
+g_1(q^2)(\epsilon^*\!\cdot v_i)v_f^\mu
+g_2(q^2)(\epsilon^*\!\cdot v_i)v_i^\mu\right].
\label{eq:scalar-axial-general}
\end{equation}
At zero recoil, $v_i=v_f$ and $\epsilon^*\!\cdot v_f=0$, so also
$\epsilon^*\!\cdot v_i=0$. In addition, the vector structure in
Eq.~\eqref{eq:scalar-vector-current} vanishes because the Levi--Civita tensor
is contracted with two identical four-velocities. Hence at exact zero recoil
only the axial structure proportional to $f(q^2)\epsilon^{*\mu}$ survives.

We therefore employ the minimal reduced left-handed current
\begin{equation}
\widehat H_S^\mu=-F_S(q^2)\epsilon^{*\mu}.
\label{eq:minimal-scalar-current}
\end{equation}
The minus sign follows from $H^\mu=(V^\mu-A^\mu)/2$; before the left-handed
factor, $\widehat A_S^\mu=2F_S(q^2)\epsilon^{*\mu}$.
Equation~\eqref{eq:minimal-scalar-current} is therefore motivated by the
zero-recoil structure of the transition. The integrated inclusive width,
however, samples the complete kinematic range in $q^2$. Away from zero recoil,
$v_i$ and $v_f$ are not parallel and the structures proportional to
$g_V(q^2)$, $g_1(q^2)$, and $g_2(q^2)$ need not vanish. Extending
Eq.~\eqref{eq:minimal-scalar-current} over the full recoil region is thus an
endpoint-motivated benchmark assumption and not a unique finite-recoil
description of the scalar-to-axial current.

The nonrelativistic spin relation fixes the relative spin-tensor structure of
the scalar-to-axial matrix element, but not the complete hadronic
normalization. We absorb into $F_S(q^2_{\rm max})$ the unknown spatial overlap,
short-distance matching effects, and convention-dependent normalization of
the composite diquark states. In particular,
\begin{equation*}
\bar c\gamma^i\gamma_5 b\longrightarrow\chi_c^\dagger\sigma^i\chi_b,
\end{equation*}
and Eq.~\eqref{eq:scalar-spin-sign} establishes the spin selection rule and
relative spin coefficient, not a
symmetry-protected normalization of the full form factor.

We therefore factorize the unknown normalization from the normalized recoil
shape according to
\begin{equation}
F_S(q^2)=F_S(q^2_{\rm max})\,\widehat F_S(q^2)
\equiv\kappa_S\widehat F_S(q^2),\qquad
\widehat F_S(q^2)=\frac{I_Y(k(q^2)/2)}{I_Y(0)},
\label{eq:scalar-normalization-shape}
\end{equation}
where
\begin{equation*}
I_Y(k)=\int dr\,u_{cc}(r)u_{bc}(r)j_0(kr).
\end{equation*}
The denominator $I_Y(0)$ denotes vanishing recoil momentum $k=0$, which
corresponds to $q^2=q^2_{\rm max}$; it is not the value at $q^2=0$. Hence
$\widehat F_S(q^2_{\rm max})=1$. The argument $k/2$ follows the recoil-sharing
convention in Yang's equal-mass final-$cc$ system. This prescription imports
only the normalized recoil profile: neither Ridgway--Wise nor Yang provides
the complete scalar-to-axial form factors required here
\cite{RidgwayWise:2019,Yang:2026}. It is important to distinguish the two normalizations in
Eq.~\eqref{eq:scalar-normalization-shape}. The condition
$\widehat F_S(q^2_{\rm max})=1$ follows by definition of the normalized shape;
it does not imply the physical normalization $F_S(q^2_{\rm max})=1$. The
latter corresponds to the central benchmark choice
\begin{equation*}
\kappa_S=F_S(q^2_{\rm max})=1,\qquad\delta_S=0,
\end{equation*}
which is a model choice rather than a symmetry-protected normalization or a
dynamical determination of the relative phase.

For this central benchmark the scalar-to-axial contribution is comparable to,
and somewhat larger than, the unmixed axial contribution. This numerical
comparison is conditional on the benchmark normalization and is not a
first-principles prediction of the scalar-channel strength.

Although Eq.~\eqref{eq:rephasing-invariance} removes arbitrary basis phases,
the dynamical relative phase between the Qin axial amplitude and the modeled
scalar amplitude is not fixed because they are not calculated in a common
microscopic framework. We therefore parameterize the scalar amplitude as
\begin{equation*}
\mathcal M_S\to\kappa_S e^{i\delta_S}\mathcal M_S^{(0)}.
\end{equation*}
For the real reference interference used here, this gives
\begin{equation*}
\Gamma_{SS}\to\kappa_S^2\Gamma_{SS},\qquad
\Gamma_{AS}\to\kappa_S\cos\delta_S\,\Gamma_{AS}.
\end{equation*}
The range $0.8\le\kappa_S\le1.2$ used below fixes $\delta_S=0$ and is a
representative local
sensitivity range, not a statistically derived uncertainty and not a bound on
the full theoretical uncertainty of $F_S(q^2_{\rm max})$.

A complete calculation would require the form factors
$f(q^2),g_1(q^2),g_2(q^2),g_V(q^2)$ in
Eqs.~\eqref{eq:scalar-vector-current} and \eqref{eq:scalar-axial-general} from
a microscopic model or effective-theory matching calculation. Away from zero
recoil the omitted structures need not vanish. Their omission is therefore a
genuine model assumption affecting absolute scalar widths and spin-resolved
interference. A rescaling of $\kappa_S$ changes the strength of the retained
$\epsilon^{*\mu}$ structure but cannot reproduce the independent recoil and
helicity dependence of the omitted Lorentz structures. By contrast, the
spin-summed cancellation derived next follows from the baryon spin algebra
under the stated assumptions.

\section{Baryon spin recoupling and final-spin decomposition}
\label{sec:spin}

The physical baryons contain a light quark with spin $s_q=1/2$. For the scalar
configuration,
\begin{equation*}
|S;J_i=\tfrac12,M_i\rangle=|S_{bc}=0,m=0\rangle|s_q=M_i\rangle.
\end{equation*}
For the axial configuration,
\begin{equation*}
|A;J_i=\tfrac12,M_i\rangle=\sum_{m_i,s}
C^{1/2,M_i}_{1,m_i;\,1/2,s}|1,m_i\rangle|s\rangle.
\end{equation*}
Using standard Clebsch--Gordan phases~\cite{Edmonds:1957,Rose:1957}, for example,
\begin{equation*}
|A;\tfrac12,\tfrac12\rangle=\sqrt{\frac23}|1,1\rangle|\downarrow_q\rangle
-\sqrt{\frac13}|1,0\rangle|\uparrow_q\rangle,
\end{equation*}
where $|1,m\rangle\equiv|A_{bc},m\rangle$ is defined in
Eq.~\eqref{eq:ordered-bc-spin-basis}.
The corresponding scalar state is
\begin{equation*}
|S;\tfrac12,\tfrac12\rangle=\frac1{\sqrt2}
(|\uparrow_b\downarrow_c\uparrow_q\rangle-|\downarrow_b\uparrow_c\uparrow_q\rangle).
\end{equation*}
These states are orthogonal.

As an internal normalization check, averaging the physical $J_i=1/2$
axial-basis baryon over $M_i=\pm1/2$ and tracing over the spectator light-quark
spin gives a completely unpolarized heavy-diquark density matrix,
\begin{equation*}
\rho_A=\frac13\mathbf 1_{3\times3}.
\end{equation*}
Thus the baryon-level recoupling reproduces the $1/3$ initial axial-diquark
polarization average of the Qin baseline rather than introducing an extra
spin normalization.

The final $cc$ diquark has $S_{cc}=1$, and
\begin{equation*}
1\otimes\frac12=\frac12\oplus\frac32.
\end{equation*}
Denoting the corresponding projectors by $P_{1/2}$ and $P_{3/2}$,
\begin{equation}
P_{1/2}+P_{3/2}=1.
\label{eq:projector-completeness}
\end{equation}

The baryon-level reduced amplitudes are
\begin{equation}
H_A^\mu(J_f,M_f;M_i)=\sum_{m_f,m_i,s}
C^{J_f,M_f}_{1,m_f;\,1/2,s}
C^{1/2,M_i}_{1,m_i;\,1/2,s}
 h_A^\mu(m_f,m_i),
\label{eq:baryon-amplitude-A}
\end{equation}
and
\begin{equation}
H_S^\mu(J_f,M_f;M_i)=\sum_{m_f,s}
C^{J_f,M_f}_{1,m_f;\,1/2,s}\delta_{s,M_i}h_S^\mu(m_f),
\label{eq:baryon-amplitude-S}
\end{equation}
with the common $\sqrt{2M_iM_f}$ stripped. At baryon-tensor level the common
factor is restored, while the spin average becomes $\frac12\sum_{M_i}$.

For fixed $J_f$ define $\Gamma_{AA}^{J_f}$, $\Gamma_{SS}^{J_f}$, and
$\Gamma_{AS}^{J_f}$ from the corresponding spin-summed quadratic and
interference terms. The mixed widths are
\begin{equation}
\Gamma_1^{J_f}=\cos^2\theta\,\Gamma_{AA}^{J_f}
+\sin^2\theta\,\Gamma_{SS}^{J_f}
-2\sin\theta\cos\theta\,\Gamma_{AS}^{J_f},
\label{eq:mixed-width-1}
\end{equation}
and
\begin{equation}
\Gamma_2^{J_f}=\sin^2\theta\,\Gamma_{AA}^{J_f}
+\cos^2\theta\,\Gamma_{SS}^{J_f}
+2\sin\theta\cos\theta\,\Gamma_{AS}^{J_f}.
\label{eq:mixed-width-2}
\end{equation}
The orthogonality of the mixing rotation gives the useful cross-check
\begin{equation*}
\Gamma_1^{J_f}+\Gamma_2^{J_f}=\Gamma_{AA}^{J_f}+\Gamma_{SS}^{J_f}
\end{equation*}
for $\kappa_S=1$, or $\Gamma_{AA}^{J_f}+\kappa_S^2\Gamma_{SS}^{J_f}$ for
general $\kappa_S$.

Using Clebsch--Gordan completeness,
\begin{equation*}
\sum_{J_f,M_f}C^{J_f,M_f}_{1,m_f;\,1/2,s}
C^{J_f,M_f*}_{1,m_f';\,1/2,s'}=\delta_{m_fm_f'}\delta_{ss'},
\end{equation*}
the total interference tensor reduces to
\begin{equation*}
W_{AS,\rm tot}^{\mu\nu}=\frac12\sum_{M_i}\sum_{m_f,m_i,s}
C^{1/2,M_i}_{1,m_i;\,1/2,s}\delta_{s,M_i}
 h_A^\mu(m_f,m_i)h_S^{\nu*}(m_f).
\end{equation*}
The delta sets $s=M_i$, and angular-momentum conservation then requires
$m_i=0$. Hence
\begin{equation*}
W_{AS,\rm tot}^{\mu\nu}\propto\frac12\sum_{M_i=\pm1/2}
C^{1/2,M_i}_{1,0;\,1/2,M_i}\sum_{m_f}h_A^\mu(m_f,0)h_S^{\nu*}(m_f).
\end{equation*}
With our phase convention,
\begin{equation*}
C^{1/2,+1/2}_{1,0;\,1/2,+1/2}=-\sqrt{\frac13},\qquad
C^{1/2,-1/2}_{1,0;\,1/2,-1/2}=+\sqrt{\frac13},
\end{equation*}
so the initial-spin average cancels exactly:
\begin{equation}
W_{AS,\rm tot}^{\mu\nu}=0,\qquad \Gamma_{AS}^{\rm tot}=0.
\label{eq:AS-total-zero}
\end{equation}
Therefore
\begin{equation}
\Gamma_{AS}^{1/2}=-\Gamma_{AS}^{3/2}.
\label{eq:AS-sector-cancel}
\end{equation}
For $c=\cos\theta$ and $s=\sin\theta$, it follows that the spin-summed mixed
widths contain no scalar--axial interference term:
\begin{align}
\Gamma_1^{\rm tot}&=c^2\Gamma_{AA}^{\rm tot}
+s^2\kappa_S^2\Gamma_{SS}^{\rm tot},\nonumber\\
\Gamma_2^{\rm tot}&=s^2\Gamma_{AA}^{\rm tot}
+c^2\kappa_S^2\Gamma_{SS}^{\rm tot}.
\label{eq:mixed-total-no-interference}
\end{align}
Thus the interference redistributes decay strength between $J_f=1/2$ and
$J_f=3/2$ without changing their sum. The equal-and-opposite relation is the
structural result of the adopted spin algebra, whereas the numerical amount
of redistributed strength depends on $\kappa_S$ and on the scalar-current
model.
This cancellation is exact within the adopted setup: the initial baryon is
unpolarized, the light quark is a spectator, the same light-cloud and spatial
configuration is used for both final-spin partners, both $J_f=1/2$ and $3/2$
are included, and spin-dependent phase-space splittings inside this multiplet
are neglected. Corrections beyond these assumptions can generate a residual
summed interference.

\section{Numerical results}
\label{sec:results}

We first display the basis-sector components because they are independent of
the choice of physical mixing angle. Unless otherwise stated, the scalar
benchmark uses $\kappa_S=1$ and $\delta_S=0$. The decomposition is shown in
Table~\ref{tab:spin-sector-widths}.

\begin{table}[t]
\centering
\caption{Angle-independent spin-sector components of the inclusive widths.
All widths are in GeV.}
\label{tab:spin-sector-widths}
\begin{tabular}{c|ccc|c}
\toprule
Final sector&$\Gamma_{AA}$&$\Gamma_{SS}$&$\Gamma_{AS}$&$\eta_{J_f}$\\
\midrule
$J_f=1/2$&$1.38762\times10^{-13}$&$8.65669\times10^{-14}$&$-8.82571\times10^{-14}$&$-0.80526$\\
$J_f=3/2$&$4.76317\times10^{-14}$&$1.73134\times10^{-13}$&$+8.82571\times10^{-14}$&$+0.97188$\\
Total&$1.86394\times10^{-13}$&$2.59701\times10^{-13}$&$-2.87\times10^{-30}$&$\simeq0$\\
\bottomrule
\end{tabular}
\end{table}

We define
\begin{equation*}
\eta_{J_f}=\frac{\Gamma_{AS}^{J_f}}
{\sqrt{\Gamma_{AA}^{J_f}\Gamma_{SS}^{J_f}}}.
\end{equation*}
The Cauchy--Schwarz inequality requires $|\eta_{J_f}|\le1$, and the values in
Table~\ref{tab:spin-sector-widths} satisfy this bound. The digits are retained
for code-level reproduction and should not be interpreted as the intrinsic
precision of the model. For orientation, Ref.~\cite{QinEtAl:2022} quotes
\begin{equation*}
\Gamma(\Xi_{bc}\to X_{cc})=(1.9\pm0.1\pm0.3\pm0.4)\times10^{-13}\,\GeV
\end{equation*}
from its stated input variations and further estimates uncalculated $v^2$
power corrections at the roughly $30\%$ level. These baseline uncertainties
are not propagated through our scenario tables, and no statistical
uncertainty is assigned to the minimal scalar-to-axial model.

The physical admixtures and their convention translations are collected in
Table~\ref{tab:mixing-scenarios}. The published baryon masses in the last
column are quoted only to identify the source state ordering. They are not
inserted into the Qin heavy-diquark phase space.

\begin{table}[t]
\centering
\small
\caption{Mixing scenarios and convention mappings. Here
$|S_{\rm lit}\rangle=-|S\rangle$ denotes the scalar basis used in the cited
literature. The decay calculation keeps the common Qin value
$M_i=6.141$ GeV for both states.}
\label{tab:mixing-scenarios}
\begin{tabular}{p{0.15\textwidth}p{0.32\textwidth}p{0.12\textwidth}p{0.27\textwidth}}
\toprule
Scenario&Published admixture or origin&Our angle&Published mass order\\
\midrule
QCD sum rule~\cite{AlievAziziSavci:2012}
&Rotation in $(A,S_{\rm lit})$ with $\phi=16^\circ\pm5^\circ$
&$16^\circ\pm5^\circ$
&No two-state mass assignment in the angle extraction\\[2pt]
Roberts--Pervin~\cite{RobertsPervin:2009}
&$0.92A+0.39S_{\rm lit}$ and $0.92S_{\rm lit}-0.39A$
&$22.97^\circ$
&$M_1=7.011$ GeV, $M_2=7.047$ GeV; state 1 lighter\\[2pt]
Decay diagnostic
&Minimum of the auxiliary destructive-overlap diagnostic
&$\theta_{\rm B}=41.06^\circ$
&No spectroscopy mass prediction\\[2pt]
AHN~\cite{AlbertusEtAl:2010a,AlbertusEtAl:2010b}
&$0.902S_{\rm lit}+0.431A$ and $-0.431S_{\rm lit}+0.902A$
&$64.46^\circ$
&$M_1=6.967$ GeV, $M_2=6.919$ GeV; state 2 lighter\\
\bottomrule
\end{tabular}
\end{table}

The corresponding spin fractions are defined by
\begin{equation*}
F_i(J_f)=\frac{\Gamma_i^{J_f}}{\Gamma_i^{1/2}+\Gamma_i^{3/2}},
\end{equation*}
and the convention-matched numerical comparison is given in
Table~\ref{tab:scenario-results}. The QCD-sum-rule row uses its central angle;
the quoted $\pm5^\circ$ range is not interpreted as a combined uncertainty of
the present decay model.

The spin projection performed here concerns the spectator-preserving $S$-wave
$ccq$ channel. In using the Qin inclusive normalization to quote
spin-resolved $\Xi_{cc}$ and $\Xi_{cc}^{*}$ yields, we assume that this
ground-state light-cloud channel gives the dominant contribution relevant for
the displaced $\Xi_{cc}$ signal. Fragmentation into $\Omega_{cc}$,
tetraquark channels, radial or orbital excitations, and light-quark
rearrangement is not modeled explicitly. Such contributions do not alter the
Clebsch--Gordan identity derived for the adopted $ccq$ multiplet, but they can
change the mapping of the partonic inclusive rate onto an observed $\Xi_{cc}$
sample.

Using the reference lifetime $\tau(\Xi_{bc}^+)=508$ fs gives the branching
fractions in Table~\ref{tab:scenario-results}. The same reference
lifetime and common initial mass are assigned to both mixed states; physical
hyperfine mass splittings and possible lifetime differences are neglected, so
the quoted branching fractions scale linearly with the corresponding physical
lifetimes. To retain comparability with Ref.~\cite{QinEtAl:2022}, we also apply its
phenomenological isospin/hadronization factor $1/2$ when quoting the
$\Xi_{cc}^{++}$ channel. This factor is a convention inherited from the Qin
comparison and is not a consequence of the spin algebra derived here.

\begin{table}[t]
\centering
\small
\caption{Numerical comparison of literature-motivated admixtures and the
separately labeled decay diagnostic at $\kappa_S=1$ and $\delta_S=0$.
Branching fractions are for $\Xi_{cc}^{++}+X$ in percent and include the Qin
comparison factor $1/2$. ``Inclusive'' includes the unresolved
$J_f=3/2$ feed-down contribution.}
\label{tab:scenario-results}
\begin{tabular}{lccccccc}
\toprule
Scenario&$\theta$&$F_1(1/2)$&$F_2(3/2)$&$\BR_1^{\rm direct}$&$\BR_2^{\rm direct}$&$\BR_1^{\rm incl}$&$\BR_2^{\rm incl}$\\
\midrule
QCD sum rule&$16.00^\circ$&0.946&0.828&7.007&1.689&7.408&9.807\\
Roberts--Pervin&$22.97^\circ$&0.983&0.875&7.496&1.200&7.624&9.591\\
Decay diagnostic&$41.06^\circ$&0.934&0.905&7.859&0.836&8.413&8.801\\
AHN&$64.46^\circ$&0.670&0.698&6.365&2.331&9.496&7.719\\
\bottomrule
\end{tabular}
\end{table}

One can see that the preference of state 1 for $J_f=1/2$ and of state 2 for
$J_f=3/2$ is present in all three literature-motivated rows, although its
numerical strength changes substantially. The value $\theta_{\rm B}$ is shown
only to illustrate the decay-interference region and is not used to select a
physical hyperfine scenario. The comparison also shows that the direct
branching fractions are more strongly separated than the spin-summed
feed-down-inclusive branching fractions.

\begin{figure}[t]
\centering
\includegraphics[width=0.72\textwidth]{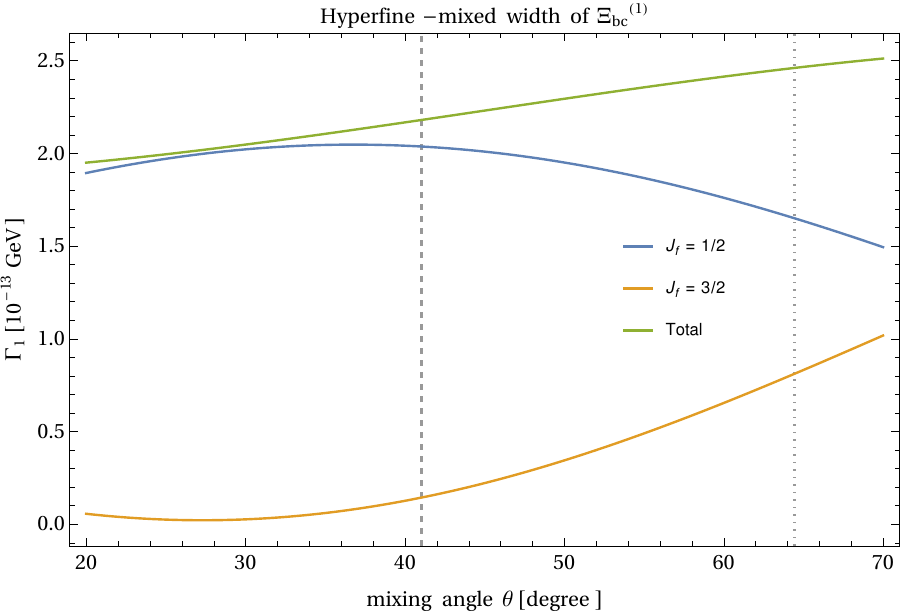}
\caption{Mixing-angle dependence of $\Gamma_1$ for $\Xi_{bc}^{(1)}$. The
dashed line marks the decay diagnostic $\theta_{\rm B}=41.06^\circ$, while
the dotted line marks the AHN literature scenario
$\theta_{\rm AHN}=64.46^\circ$.}
\label{fig:gamma1}
\end{figure}

\begin{figure}[t]
\centering
\includegraphics[width=0.72\textwidth]{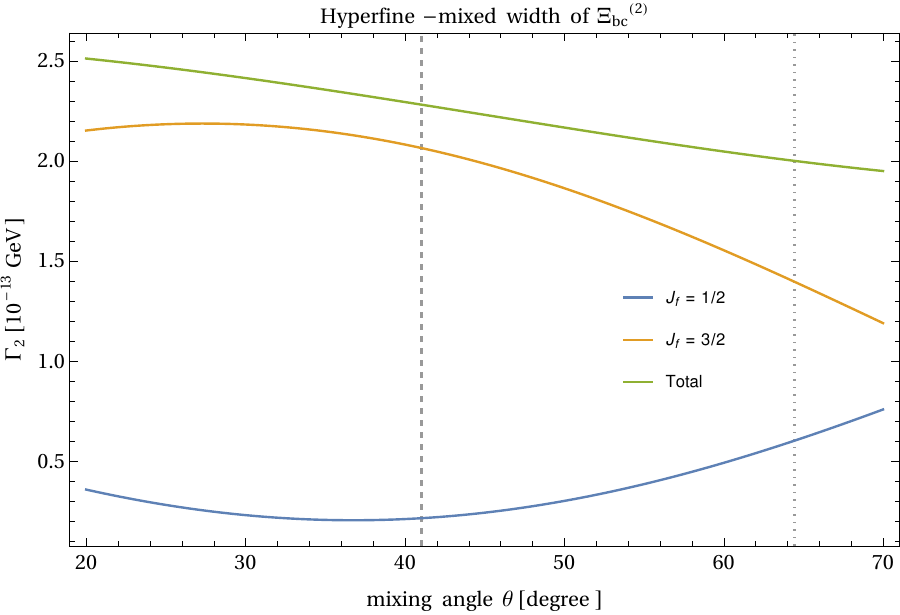}
\caption{Mixing-angle dependence of $\Gamma_2$ for $\Xi_{bc}^{(2)}$. The
dashed line marks the decay diagnostic $\theta_{\rm B}=41.06^\circ$, while
the dotted line marks the AHN literature scenario
$\theta_{\rm AHN}=64.46^\circ$. The $J_f=3/2$ contribution remains dominant
at both displayed angles.}
\label{fig:gamma2}
\end{figure}

\begin{figure}[t]
\centering
\includegraphics[width=0.94\textwidth]{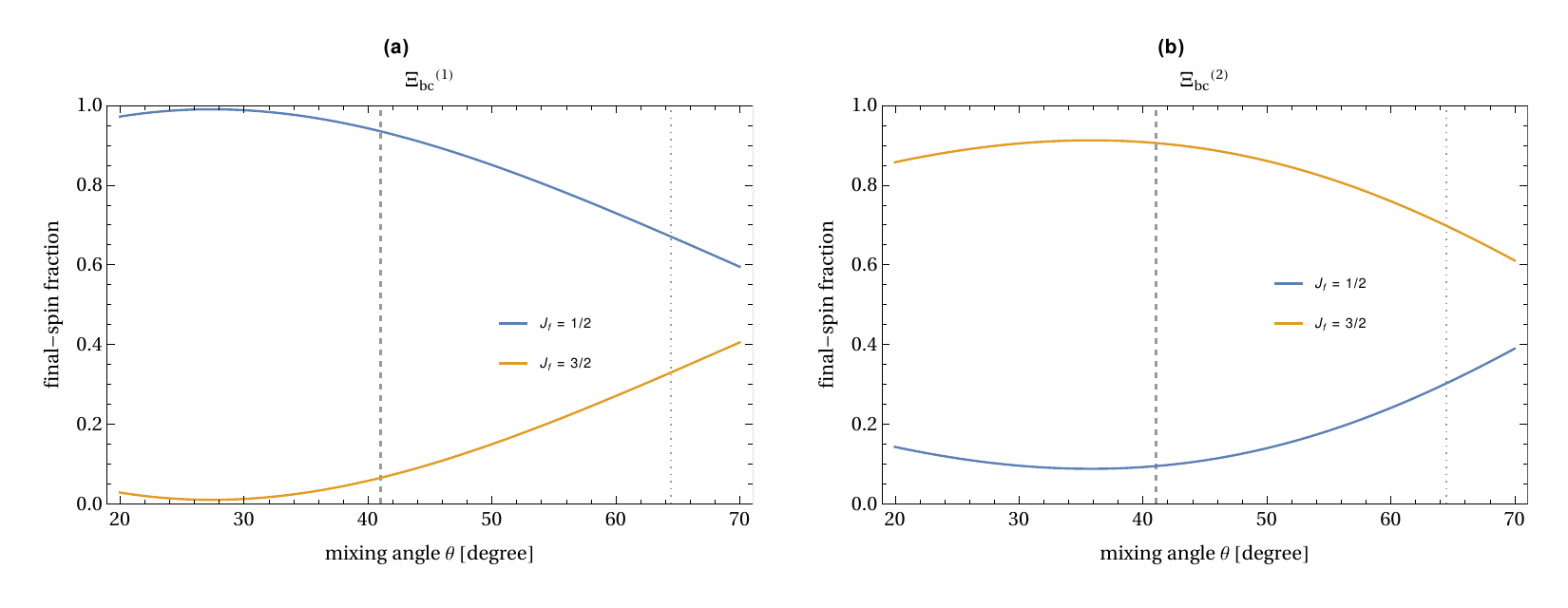}
\caption{Final-spin fractions, $F_i(J_f,\theta)$, as functions of the mixing angle for
$\Xi_{bc}^{(1)}$ and $\Xi_{bc}^{(2)}$. The dashed line marks the decay
diagnostic $\theta_{\rm B}=41.06^\circ$, and the dotted line marks the AHN
literature scenario $\theta_{\rm AHN}=64.46^\circ$.}
\label{fig:spin-fractions}
\end{figure}

\begin{figure}[t]
\centering
\includegraphics[width=0.72\textwidth]{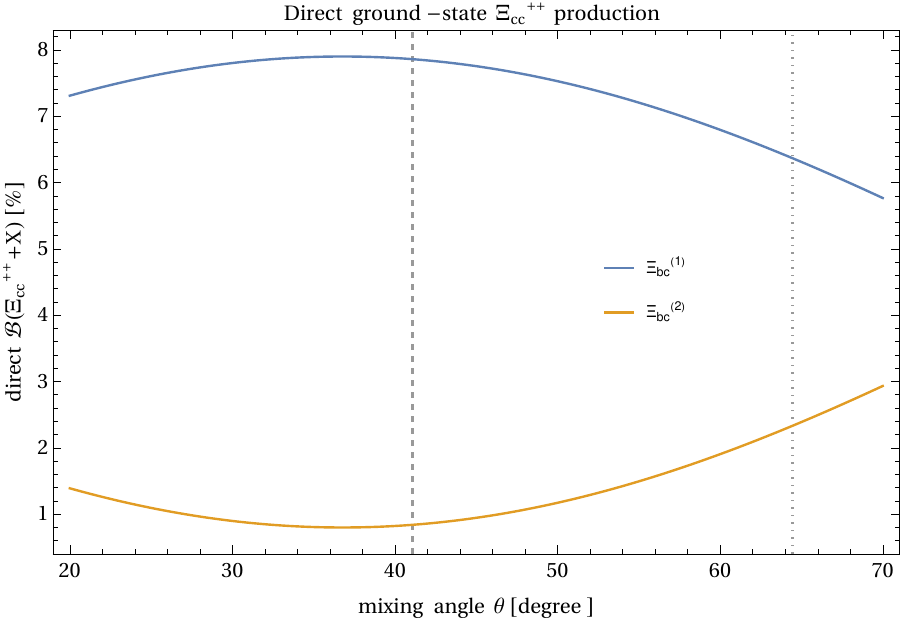}
\caption{Direct ground-state $\Xi_{cc}^{++}$ branching fractions as functions
of the hyperfine mixing angle. The dashed line marks the decay diagnostic
$\theta_{\rm B}=41.06^\circ$, and the dotted line marks the AHN literature
scenario $\theta_{\rm AHN}=64.46^\circ$.}
\label{fig:direct-br}
\end{figure}

\begin{figure}[t]
\centering
\includegraphics[width=0.72\textwidth]{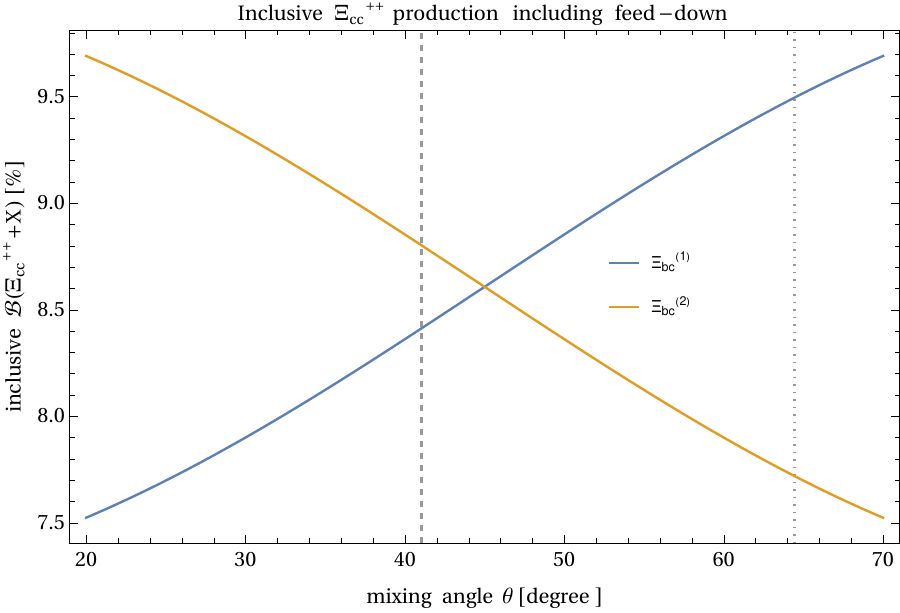}
\caption{Inclusive $\Xi_{cc}^{++}$ branching fractions including unresolved
feed-down from the $J_f=3/2$ $\Xi_{cc}^{*}$ sector.}
\label{fig:inclusive-br}
\end{figure}

Figures~\ref{fig:gamma1}--\ref{fig:inclusive-br} show the mixing-angle
dependence over a range containing all scenarios in
Table~\ref{tab:mixing-scenarios}. The two vertical lines retained in the
figures identify the decay diagnostic and AHN points; the QCD-sum-rule and
Roberts--Pervin values are reported numerically in
Table~\ref{tab:scenario-results}. The spin selectivity is not restricted to
the diagnostic point, although its magnitude depends on the scalar
normalization as discussed in Sec.~\ref{sec:robustness}.

\FloatBarrier
\section{Dependence on Model Inputs}
\label{sec:robustness}

\subsection{Mixing-Angle and Scalar-Normalization Sensitivity}

For general $\kappa_S$ and $\delta_S$,
\begin{align*}
\Gamma_1^{J_f}(\theta,\kappa_S,\delta_S)&=\cos^2\theta\,\Gamma_{AA}^{J_f}
+\sin^2\theta\,\kappa_S^2\Gamma_{SS}^{J_f}
-2\sin\theta\cos\theta\,\kappa_S\cos\delta_S\,\Gamma_{AS}^{J_f},\\
\Gamma_2^{J_f}(\theta,\kappa_S,\delta_S)&=\sin^2\theta\,\Gamma_{AA}^{J_f}
+\cos^2\theta\,\kappa_S^2\Gamma_{SS}^{J_f}
+2\sin\theta\cos\theta\,\kappa_S\cos\delta_S\,\Gamma_{AS}^{J_f}.
\end{align*}
The spin-summed interference still cancels because
$\Gamma_{AS}^{1/2}=-\Gamma_{AS}^{3/2}$.

We scan $35^\circ\le\theta\le50^\circ$ and
$0.8\le\kappa_S\le1.2$. This is deliberately a local sensitivity test around
the decay-diagnostic region at fixed $\delta_S=0$. It is neither a physical
hyperfine-angle uncertainty band nor a confidence interval, and it is not a
complete theoretical uncertainty band. The results are summarized in
Table~\ref{tab:robustness-scan}.

\begin{table}[t]
\centering
\caption{Local diagnostic sensitivity scan over
$35^\circ\le\theta\le50^\circ$ and $0.8\le\kappa_S\le1.2$. This interval is
not a physical mixing-angle error band.}
\label{tab:robustness-scan}
\begin{tabular}{lcc}
\toprule
Quantity&Minimum&Maximum\\
\midrule
$\BR_1^{\rm incl}$ [\%]&6.73572&11.4405\\
$\BR_2^{\rm incl}$ [\%]&6.67015&12.0499\\
$\BR_1^{\rm direct}$ [\%]&6.15030&9.17855\\
$\BR_2^{\rm direct}$ [\%]&0.635934&1.34241\\
$F_1(J_f=1/2)$&0.791938&0.988497\\
$F_2(J_f=3/2)$&0.804627&0.911849\\
$\BR_1^{\rm direct}/\BR_2^{\rm direct}$&4.58154&10.7926\\
$\BR_1^{\rm incl}/\BR_2^{\rm incl}$&0.794547&1.12343\\
\bottomrule
\end{tabular}
\end{table}

Because $F_S(q^2_{\rm max})$ is not fixed from first principles, the pronounced
complementary spin redistribution must be interpreted conditionally. In the
formal limit $\kappa_S\to0$ the scalar channel decouples and both mixed states
inherit the axial-channel fractions
\begin{equation*}
F_A(J_f=\tfrac12)=\frac{\Gamma_{AA}^{1/2}}
{\Gamma_{AA}^{1/2}+\Gamma_{AA}^{3/2}}\simeq0.744,
\qquad F_A(J_f=\tfrac32)\simeq0.256.
\end{equation*}
The strong complementary pattern then disappears. In the opposite formal
limit $\kappa_S\to\infty$, both states approach the pure scalar fractions
\begin{equation*}
F_S(J_f=\tfrac12)\simeq\frac13,\qquad
F_S(J_f=\tfrac32)\simeq\frac23.
\end{equation*}
Thus the large state-to-state spin selectivity is an interference phenomenon
that is strongest when axial and scalar amplitudes are comparable. The more
model-independent result of the present analysis is the spin-recoupling
identity $\Gamma_{AS}^{1/2}+\Gamma_{AS}^{3/2}=0$ under the assumptions of
Sec.~\ref{sec:spin}.

\subsection{Relative-phase dependence}

The decay-diagnostic illustration takes $\delta_S=0$. To display separately the
effect of the undetermined relative phase, Table~\ref{tab:phase-scan} compares
$\delta_S=0$, $\pi/2$, and $\pi$ at $\theta=\theta_{\rm B}=41.06^\circ$ and
$\kappa_S=1$.
The entries follow directly from the spin-sector components in
Table~\ref{tab:spin-sector-widths}.

\begin{table}[t]
\centering
\caption{Relative-phase sensitivity of the final-spin fractions at the
decay-diagnostic point $\theta=\theta_{\rm B}=41.06^\circ$ and $\kappa_S=1$.}
\label{tab:phase-scan}
\begin{tabular}{c|cccc}
\toprule
$\delta_S$&$F_1(1/2)$&$F_1(3/2)$&$F_2(1/2)$&$F_2(3/2)$\\
\midrule
$0$&0.934&0.066&0.095&0.905\\
$\pi/2$&0.533&0.467&0.478&0.522\\
$\pi$&0.132&0.868&0.862&0.138\\
\bottomrule
\end{tabular}
\end{table}

For $\delta_S=\pi/2$ the interference vanishes and both final-spin
distributions become comparatively balanced. For $\delta_S=\pi$ the
interference reverses sign, so the preferred final-spin sectors of the two
mixed states are reversed relative to the $\delta_S=0$ diagnostic. The
spin-summed widths remain
\begin{equation*}
\Gamma_1^{\rm tot}=2.18018\times10^{-13}\ \GeV,\qquad
\Gamma_2^{\rm tot}=2.28076\times10^{-13}\ \GeV
\end{equation*}
for all three phases because the sector interference terms cancel. More
generally,
\begin{equation*}
\Gamma_{AS}^{1/2}(\delta_S)+\Gamma_{AS}^{3/2}(\delta_S)
=\cos\delta_S\left(\Gamma_{AS}^{1/2}+\Gamma_{AS}^{3/2}\right)=0.
\end{equation*}
Thus the cancellation identity is independent of the common relative phase,
whereas the numerical direction and strength of the spin redistribution are
phase dependent.

\subsection{Numerical Sensitivity to the \(a_2\) Sign}
\label{sec:a2-sign-test}

It is important to mention here that changing the sign printed for
$c_{a_2}$ is a nontrivial intervention in the Qin numerical input, even though
the negative sign reproduces the expected zero-recoil normalization. To check
whether the main results depend on this choice, we have repeated the complete
axial and scalar--axial interference calculation using the literal published
value $c_{a_2}=+238.2$, while keeping all other inputs unchanged. The central
comparison in Table~\ref{tab:a2-sign-sensitivity} uses
$\theta=\theta_{\rm B}=41.05685^\circ$, $\kappa_S=1$, and $\delta_S=0$.

\begin{table}[t]
\centering
\small
\caption{Sensitivity to the sign of the Qin $a_2$ curvature parameter. The
first row is the unmixed spin-summed axial width. The remaining rows are
evaluated at the decay-diagnostic point. Branching fractions refer to
$\Xi_{cc}^{++}+X$ and are given in percent.}
\label{tab:a2-sign-sensitivity}
\begin{tabular}{lccc}
\toprule
Quantity&$c_{a_2}=-238.2$&$c_{a_2}=+238.2$&Relative change [\%]\\
\midrule
$\Gamma_A$ [GeV]&$1.86391\times10^{-13}$&$1.87140\times10^{-13}$&$+0.402$\\
$F_1(J_f=1/2)$&0.93418&0.93331&$-0.093$\\
$F_2(J_f=3/2)$&0.90503&0.90447&$-0.061$\\
$\BR_1^{\rm direct}$&7.8594&7.8674&$+0.102$\\
$\BR_2^{\rm direct}$&0.8359&0.8420&$+0.729$\\
$\BR_1^{\rm incl}$&8.4132&8.4296&$+0.196$\\
$\BR_2^{\rm incl}$&8.8013&8.8138&$+0.142$\\
\bottomrule
\end{tabular}
\end{table}

One can see that the total axial width changes by only $0.40\%$. The two
dominant final-spin fractions change by less than $0.10\%$, and the largest
variation among the quoted branching fractions is approximately $0.73\%$.
Furthermore, the spin-recoupling relation
\begin{equation*}
\Gamma_{AS}^{1/2}+\Gamma_{AS}^{3/2}=0
\end{equation*}
is unchanged because it follows from the Clebsch--Gordan algebra and not from
the numerical value of $c_{a_2}$. Therefore, the main conclusions survive the
literal published positive sign. We retain $c_{a_2}=-238.2$ for the central
results because it reproduces $a_2(q^2_{\rm max})\simeq1$.

\FloatBarrier

\subsection{Behaviour over an Extended Mixing-Angle Range}

The convention-matched literature scenarios are compared directly in
Tables~\ref{tab:mixing-scenarios} and \ref{tab:scenario-results}. As a broader
orientation, with $\kappa_S=1$, $\delta_S=0$, and
$20^\circ\le\theta\le70^\circ$,
\begin{equation*}
1.97\lesssim\frac{\BR_1^{\rm direct}}{\BR_2^{\rm direct}}\lesssim9.92,
\qquad
0.78\lesssim\frac{\BR_1^{\rm incl}}{\BR_2^{\rm incl}}\lesssim1.29.
\end{equation*}
These continuous-angle ranges are sensitivity diagnostics rather than
statistical confidence intervals or predictions for the physical
hyperfine-mixing angle.

\subsection{Recoil-shape dependence}

The benchmark scalar recoil shape is
\begin{equation*}
\widehat F_S(q^2)=\frac{I_Y(k(q^2)/2)}{I_Y(0)}.
\end{equation*}
As a diagnostic alternative we use the normalized Qin RMS shape
\begin{equation*}
F_{\rm Qin}^{\rm RMS}(q^2)=
\frac{[(a_0^2(q^2)+a_1^2(q^2)+a_2^2(q^2))/3]^{1/2}}
{[(a_0^2(q^2_{\rm max})+a_1^2(q^2_{\rm max})+a_2^2(q^2_{\rm max}))/3]^{1/2}}.
\end{equation*}
This RMS combination is used only as a dimensionless diagnostic of the typical
recoil variation in the Qin axial-to-axial form factors; it is not identified
with a physical scalar-to-axial form factor. Both shapes are normalized to
unity at zero recoil. The integrated scalar width changes by less than one
percent under this shape replacement.

\begin{figure}[t]
\centering
\includegraphics[width=0.94\textwidth]{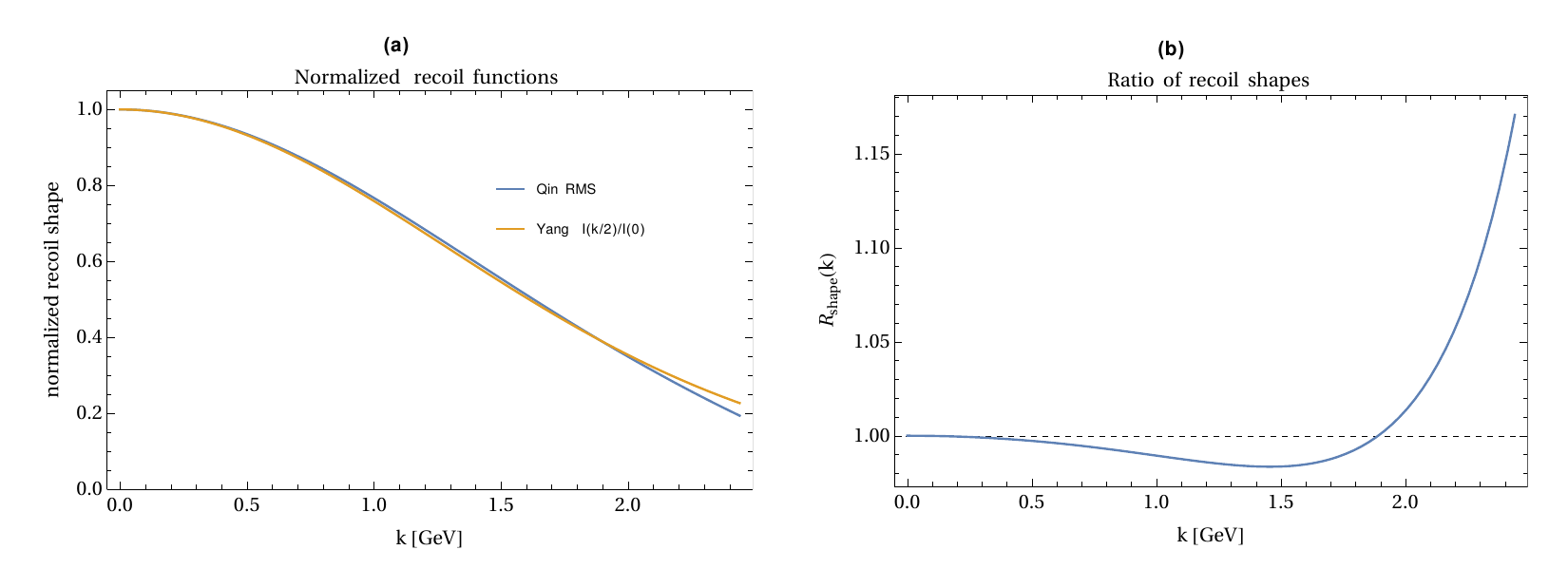}
\caption{Normalized recoil-shape comparison. Panel (a) compares the Qin RMS, $F_{Qin}^{RMS}(q^2(k^2))$,
diagnostic with the potential-model overlap proxy $\widehat F_S(q^2)$; panel (b)
shows their ratio.}
\label{fig:recoil-shape}
\end{figure}

Likewise, in an unmixed shape-only test, replacing the normalized Qin recoil
dependence by the explicit potential-model overlap while retaining the Qin
overall normalization changes the integrated width by less than one percent.
This does not imply agreement of the absolute model predictions: Yang's full
potential-model calculation gives approximately $4.1\times10^{-13}$ GeV
\cite{Yang:2026}, while the Qin baseline used here is approximately
$1.9\times10^{-13}$ GeV. The spin-sector interference cancellation is
unchanged because it follows from the Clebsch--Gordan algebra rather than the
choice of normalized recoil profile.

\FloatBarrier
\section{Discussion and phenomenological implications}
\label{sec:discussion}

Here ``inclusive'' refers to inclusiveness over the partonic system $X$; the
final doubly charmed baryon is restricted to the two members of the adopted
ground-state $ccq$ spin multiplet. The scalar--axial interference is large and
opposite in the two spin sectors but cancels in their sum. Consequently, for
the benchmark scalar normalization the convention-matched QCD-sum-rule,
Roberts--Pervin, and AHN admixtures all show a stronger separation in the
direct $\Xi_{cc}^{++}$ contributions than in the corresponding spin-summed
feed-down-inclusive branching fractions. The numerical size of this effect
varies across the three scenarios.

This conclusion has to be interpreted at two different levels. The
cancellation
\begin{equation*}
\Gamma_{AS}^{1/2}+\Gamma_{AS}^{3/2}=0
\end{equation*}
is a spin-recoupling result within the stated spectator/common-multiplet
assumptions and is invariant under a consistent rephasing of the scalar basis.
By contrast, the magnitude and direction of the complementary final-spin
redistribution depend on the size and dynamical relative phase of the
scalar-to-axial amplitude. Provided that the scalar-to-axial transition is
not strongly suppressed relative to the axial-to-axial amplitude and the
relative phase permits substantial interference, hyperfine mixing can produce
a pronounced redistribution between $J_f=1/2$ and $J_f=3/2$.

The value $\theta_{\rm B}=41.06^\circ$ has a different status from the
literature admixtures. It is obtained by minimizing an auxiliary destructive
decay overlap and is retained only as a diagnostic of the interference region.
It is not a prediction from a hyperfine mass Hamiltonian. Moreover, the
literature calculations do not agree on the physical state ordering: the
Roberts--Pervin masses place state 1 below state 2, whereas the AHN masses
place their state 1 above state 2. We quote these masses to make the state
labels explicit, but use a common Qin heavy-diquark mass in all decay rows so
that Table~\ref{tab:scenario-results} isolates the admixture dependence.

The experimental meaning of the inclusive rate also depends on feed-down. If
$\Xi_{cc}^{*}$ decays into the corresponding ground-state $\Xi_{cc}$ plus soft
electromagnetic radiation or, if kinematically allowed, soft hadrons, and the
soft products are not used to separate production categories, the observed
$\Xi_{cc}^{++}$ sample can contain the $J_f=3/2$ component. The feed-down
interpretation therefore assumes an effectively dominant cascade of the
spin-excited partner to the ground-state channel. If direct production can be
separated from feed-down, the final-spin composition becomes much more
sensitive to the hyperfine-mixing pattern.

Actual event yields will additionally depend on the production rates of the
two physical $\Xi_{bc}$ states, their lifetimes, detector efficiencies, and
acceptance for feed-down topologies. These effects lie beyond the present
calculation. Likewise, within the scope of this calculation the spin-summed
inclusive prediction contains contributions from both spin members of the
adopted ground-state $ccq$ multiplet; it should not be interpreted as an
explicit calculation of all possible doubly charmed hadronic configurations.

A complete first-principles calculation of the scalar-to-axial form factors
would reduce the main model dependence. Physical $\Xi_{cc}$--$\Xi_{cc}^{*}$
mass splittings, spin-dependent spatial wave functions, corrections to the
spectator approximation, perturbative QCD corrections, and nonperturbative
power corrections can all modify absolute rates and can generate a residual
spin-summed interference. Nevertheless, the distinction between a relatively
protected spin-summed interference cancellation and a conditional but
potentially large spin redistribution is a useful organizing principle for
future phenomenology.

\section{Summary}
\label{sec:summary}

We have studied hyperfine mixing in the inclusive transition
$\Xi_{bc}\to H_{cc}+X$ by combining the Qin heavy-diquark baseline with an
explicit baryon-spin recoupling and a minimal scalar-to-axial transition
model. The physical states are written as orthogonal mixtures of
$S_{bc}=1$ and $S_{bc}=0$ components, while the final $S_{cc}=1$ diquark
couples to the spectator light quark to form the $J_f=1/2$ and $J_f=3/2$
ground-state spin partners.

Previous studies established that hyperfine mixing can strongly redistribute
exclusive decay strength between the two final-spin channels. In the present
work, the new element is its incorporation into the Qin inclusive framework,
together with the explicit scalar--axial interference analysis and the
distinction between the direct and feed-down-inclusive contributions.

The central structural result is
\begin{equation*}
\Gamma_{AS}^{1/2}=-\Gamma_{AS}^{3/2},\qquad \Gamma_{AS}^{\rm tot}=0.
\end{equation*}
It follows from Clebsch--Gordan completeness and the spin structure of the
initial scalar and axial configurations, under the stated unpolarized,
spectator, common-multiplet, and equal-phase-space assumptions.

For the benchmark choice $F_S(q^2_{\rm max})=1$, the scalar-to-axial and
axial-to-axial amplitudes are comparable and their interference produces a
complementary final-spin pattern across the convention-matched QCD-sum-rule,
Roberts--Pervin, and AHN scenarios: $\Xi_{bc}^{(1)}$ preferentially feeds
$J_f=1/2$, while $\Xi_{bc}^{(2)}$ preferentially feeds $J_f=3/2$. The direct
ground-state branching fractions are therefore more sensitive to the mixing
pattern than the spin-summed inclusive branching fractions including
unresolved feed-down. The value $\theta_{\rm B}=41.06^\circ$ is retained only
as a decay-interference diagnostic and is not a prediction of the physical
hyperfine angle. This numerical selectivity is conditional on the
magnitude and relative phase of the scalar-to-axial amplitude. It is lost in
the limits $\kappa_S\to0$ or $\kappa_S\to\infty$, becomes weak when
$\delta_S=\pi/2$, and reverses direction when $\delta_S=\pi$.

The normalized recoil-shape tests show that the interference cancellation and
the displayed spin redistribution are not artifacts of the particular recoil
profile used. However, the absolute branching fractions remain model
dependent because the complete scalar-to-axial form factors are not known.
A separate use of the literal published value $c_{a_2}=+238.2$ changes the
quoted branching fractions by less than $0.8\%$ and leaves the main conclusions
unchanged.
A dedicated calculation from NRQCD, potential models, lattice QCD, or
heavy-diquark effective theory would therefore be an important next step.

\section*{Data Availability Statement}
No new experimental data were generated in this work. The numerical results
are obtained from the analytic expressions and input parameters specified in
the manuscript.

\appendix
\renewcommand{\theequation}{\Alph{section}.\arabic{equation}}

\section{Kinematics and tensor normalization}
\label{app:kinematics}

The initial and final heavy-diquark momenta are
\begin{equation*}
P_i^\mu=M_i v_i^\mu,\qquad P_f^\mu=M_fv_f^\mu,
\qquad v_i^2=v_f^2=1,
\end{equation*}
with $q^\mu=P_i^\mu-P_f^\mu$. The recoil variable is
\begin{equation}
w=v_i\cdot v_f=\frac{M_i^2+M_f^2-q^2}{2M_iM_f}.
\label{eq:app-w}
\end{equation}
In the initial rest frame,
\begin{equation*}
v_i^\mu=(1,0,0,0),\qquad
v_f^\mu=\left(\frac{E_f}{M_f},0,0,\frac{k}{M_f}\right)
=(w,0,0,\sqrt{w^2-1}),
\end{equation*}
where
\begin{equation}
E_f=\frac{M_i^2+M_f^2-q^2}{2M_i},\qquad
k=\frac{\sqrt{\lambda(M_i^2,M_f^2,q^2)}}{2M_i}.
\label{eq:app-k}
\end{equation}
Zero recoil is $q^2=q^2_{\rm max}$, $k=0$, $w=1$.

The reduced axial-to-axial current is
$\widehat H_A^\mu=(\widehat V_A^\mu-\widehat A_A^\mu)/2$, with the full matrix
element obtained by multiplying by $\sqrt{2M_iM_f}$. For the baryon-level
calculation the initial spin average is $\frac12\sum_{M_i=\pm1/2}$.

For fixed $q^2$,
\begin{equation}
T(q^2,m_\ell)=2M_iM_f\,\frac12\sum_{M_i,M_f}
H^\mu K_{\mu\nu}^{(\ell)}H^{\nu*}
\label{eq:app-tensor}
\end{equation}
for the physical $J_i=1/2$ baryon. In the unmixed axial-diquark baseline this
reduces to the $1/3$ polarization average in Eq.~\eqref{eq:TA-tensor}.
The differential rate is Eq.~\eqref{eq:sl-width}, with integration limits
$m_\ell^2$ and $(M_i-M_f)^2$. The hadronic contribution is obtained with
Eq.~\eqref{eq:hadronic-replacement}.

\section{Clebsch--Gordan algebra for the interference cancellation}
\label{app:cg}

The scalar initial state is
\begin{equation*}
|S;J_i=\tfrac12,M_i\rangle=|0,0\rangle_{bc}|M_i\rangle_q,
\end{equation*}
while
\begin{equation*}
|A;J_i=\tfrac12,M_i\rangle=\sum_{m_i,s}
C^{1/2,M_i}_{1,m_i;\,1/2,s}|1,m_i\rangle_{bc}|s\rangle_q.
\end{equation*}
The final state is
\begin{equation*}
|J_f,M_f\rangle=\sum_{m_f,s}C^{J_f,M_f}_{1,m_f;\,1/2,s}
|1,m_f\rangle_{cc}|s\rangle_q.
\end{equation*}
Substitution of Eqs.~\eqref{eq:baryon-amplitude-A} and
\eqref{eq:baryon-amplitude-S} into the summed interference tensor gives
\begin{align*}
W_{AS,\rm tot}^{\mu\nu}
&=\frac12\sum_{M_i}\sum_{J_f,M_f}
\sum_{m_f,m_i,s}\sum_{m_f',s'}
C^{J_f,M_f}_{1,m_f;\,1/2,s}
C^{1/2,M_i}_{1,m_i;\,1/2,s}\\
&\qquad\times C^{J_f,M_f*}_{1,m_f';\,1/2,s'}
\delta_{s',M_i}h_A^\mu(m_f,m_i)h_S^{\nu*}(m_f').
\end{align*}
Using completeness over $J_f=1/2,3/2$ gives
\begin{equation*}
W_{AS,\rm tot}^{\mu\nu}=\frac12\sum_{M_i}
\sum_{m_f,m_i,s}C^{1/2,M_i}_{1,m_i;\,1/2,s}
\delta_{s,M_i}h_A^\mu(m_f,m_i)h_S^{\nu*}(m_f).
\end{equation*}
The delta sets $s=M_i$ and therefore $m_i=0$. The two nonzero Clebsch--Gordan
coefficients have equal magnitude and opposite sign, yielding
\begin{equation*}
-\sqrt{\frac13}+\sqrt{\frac13}=0.
\end{equation*}
Thus $W_{AS,\rm tot}^{\mu\nu}=0$ and
$\Gamma_{AS}^{1/2}=-\Gamma_{AS}^{3/2}$.

\section{Origin of the decay-interference benchmark}
\label{app:benchmark}

The auxiliary overlap-level diagnostic is
\begin{equation}
I_{\Xi_2}^{\rm eff}(p,\theta)=I_1(p/2)\sin\theta-
\frac{\sqrt3}{2}I_0(p/2)\cos\theta.
\label{eq:app-Ieff}
\end{equation}
For $I_0=I_1$, it vanishes at
\begin{equation*}
\tan\theta=\frac{\sqrt3}{2},\qquad\theta=40.89^\circ.
\end{equation*}
With the explicit spin-dependent overlap functions used in the auxiliary
calculation, the integrated diagnostic is minimized at
\begin{equation*}
\theta_{\rm min}=41.05685^\circ.
\end{equation*}
We therefore use $\theta_{\rm B}=41.06^\circ$ as a diagnostic point only;
it is not a determination of the physical hyperfine mass-mixing angle.


\begin{thebibliography}{99}
\bibitem{KiselevLikhoded:2002} V.~V.~Kiselev and A.~K.~Likhoded, Phys. Usp. \textbf{45}, 455 (2002), arXiv:hep-ph/0103169.
\bibitem{EbertFaustovGalkin:2002} D.~Ebert, R.~N.~Faustov, and V.~O.~Galkin, Phys. Rev. D \textbf{66}, 014008 (2002), arXiv:hep-ph/0201217.
\bibitem{GershteinEtAl:2000} S.~S.~Gershtein, V.~V.~Kiselev, A.~K.~Likhoded, and A.~I.~Onishchenko, Phys. Rev. D \textbf{62}, 054021 (2000), arXiv:hep-ph/9811212.
\bibitem{RobertsPervin:2008} W.~Roberts and M.~Pervin, Int. J. Mod. Phys. A \textbf{23}, 2817 (2008), arXiv:0711.2492 [nucl-th].
\bibitem{KarlinerRosner:2014} M.~Karliner and J.~L.~Rosner, Phys. Rev. D \textbf{90}, 094007 (2014), arXiv:1408.5877 [hep-ph].
\bibitem{BrownEtAl:2014} Z.~S.~Brown, W.~Detmold, S.~Meinel, and K.~Orginos, Phys. Rev. D \textbf{90}, 094507 (2014), arXiv:1409.0497 [hep-lat].
\bibitem{ShahEtAl:CPC2016} Z.~Shah, K.~Thakkar, A.~K.~Rai, and P.~C.~Vinodkumar, Chin. Phys. C \textbf{40}, 123102 (2016), arXiv:1609.08464 [nucl-th].
\bibitem{ShahEtAl:EPJA2016} Z.~Shah, K.~Thakkar, A.~K.~Rai, and P.~C.~Vinodkumar, Eur. Phys. J. A \textbf{52}, 313 (2016).
\bibitem{WengEtAl:2018} X.-Z.~Weng, W.-Z.~Chen, X.-L.~Deng, and S.-L.~Zhu, Phys. Rev. D \textbf{97}, 054008 (2018), arXiv:1801.08644 [hep-ph].
\bibitem{ShekariTousi:2024mso}
M.~Shekari Tousi and K.~Azizi,
Phys. Rev. D \textbf{109}, 054005 (2024),
arXiv:2401.07151 [hep-ph].

\bibitem{ShekariTousi:2025xox}
M.~Shekari Tousi and K.~Azizi,
Phys. Lett. B \textbf{875}, 140377 (2026),
arXiv:2508.10328 [hep-ph].

\bibitem{Tousi:2024usi}
M.~Shekari Tousi, K.~Azizi, and H.~R.~Moshfegh,
Phys. Rev. D \textbf{110}, 114001 (2024),
arXiv:2409.00241 [hep-ph].

\bibitem{ShekariTousi:2025fjf}
M.~Shekari Tousi and K.~Azizi,
Phys. Rev. D \textbf{112}, 074012 (2025),
arXiv:2504.17030 [hep-ph].
\bibitem{GeorgiWise:1990} H.~Georgi and M.~B.~Wise, Phys. Lett. B \textbf{243}, 279 (1990).
\bibitem{SavageWise:1990} M.~J.~Savage and M.~B.~Wise, Phys. Lett. B \textbf{248}, 177 (1990).
\bibitem{WhiteSavage:1991} M.~J.~White and M.~J.~Savage, Phys. Lett. B \textbf{271}, 410 (1991).
\bibitem{HuMehen:2006} J.~Hu and T.~Mehen, Phys. Rev. D \textbf{73}, 054003 (2006), arXiv:hep-ph/0511321.
\bibitem{MehenPowell:2011} T.~Mehen and J.~W.~Powell, Phys. Rev. D \textbf{84}, 114013 (2011), arXiv:1109.3479 [hep-ph].
\bibitem{IsgurWise:1989} N.~Isgur and M.~B.~Wise, Phys. Lett. B \textbf{232}, 113 (1989).
\bibitem{IsgurWise:1991} N.~Isgur and M.~B.~Wise, Nucl. Phys. B \textbf{348}, 276 (1991).
\bibitem{EichtenHill:1990} E.~Eichten and B.~Hill, Phys. Lett. B \textbf{234}, 511 (1990).
\bibitem{Grinstein:1990} B.~Grinstein, Nucl. Phys. B \textbf{339}, 253 (1990).
\bibitem{Luke:1990} M.~E.~Luke, Phys. Lett. B \textbf{252}, 447 (1990).
\bibitem{ManoharWise:1994} A.~V.~Manohar and M.~B.~Wise, Phys. Rev. D \textbf{49}, 1310 (1994), arXiv:hep-ph/9308246.
\bibitem{Neubert:1994} M.~Neubert, Phys. Rept. \textbf{245}, 259 (1994), arXiv:hep-ph/9306320.
\bibitem{ManoharWise:2000} A.~V.~Manohar and M.~B.~Wise, \emph{Heavy Quark Physics} (Cambridge University Press, 2000).
\bibitem{CaswellLepage:1986} W.~E.~Caswell and G.~P.~Lepage, Phys. Lett. B \textbf{167}, 437 (1986).
\bibitem{BodwinBraatenLepage:1995} G.~T.~Bodwin, E.~Braaten, and G.~P.~Lepage, Phys. Rev. D \textbf{51}, 1125 (1995), Erratum: Phys. Rev. D \textbf{55}, 5853 (1997), arXiv:hep-ph/9407339.
\bibitem{BrambillaEtAl:2005} N.~Brambilla, A.~Pineda, J.~Soto, and A.~Vairo, Rev. Mod. Phys. \textbf{77}, 1423 (2005), arXiv:hep-ph/0410047.
\bibitem{BrambillaEtAl:2011} N.~Brambilla \emph{et al.}, Eur. Phys. J. C \textbf{71}, 1534 (2011), arXiv:1010.5827 [hep-ph].
\bibitem{LHCb:2017XiccObservation} R.~Aaij \emph{et al.} (LHCb), Phys. Rev. Lett. \textbf{119}, 112001 (2017), arXiv:1707.01621 [hep-ex].
\bibitem{SELEX:2002Xicc} M.~Mattson \emph{et al.} (SELEX), Phys. Rev. Lett. \textbf{89}, 112001 (2002), arXiv:hep-ex/0208014.
\bibitem{SELEX:2005Xicc} A.~Ocherashvili \emph{et al.} (SELEX), Phys. Lett. B \textbf{628}, 18 (2005), arXiv:hep-ex/0406033.
\bibitem{LHCb:2018XiccDecay} R.~Aaij \emph{et al.} (LHCb), Phys. Rev. Lett. \textbf{121}, 162002 (2018), arXiv:1807.01919 [hep-ex].
\bibitem{LHCb:2018XiccLifetime} R.~Aaij \emph{et al.} (LHCb), Phys. Rev. Lett. \textbf{121}, 052002 (2018), arXiv:1806.02744 [hep-ex].
\bibitem{LHCb:2020XiccPlusSearch} R.~Aaij \emph{et al.} (LHCb), Sci. China Phys. Mech. Astron. \textbf{63}, 221062 (2020), arXiv:1909.12273 [hep-ex].
\bibitem{LHCb:2026XiccPlusObservation} R.~Aaij \emph{et al.} (LHCb), arXiv:2603.28456 [hep-ex] (2026).
\bibitem{QinEtAl:2022} Q.~Qin, Y.-J.~Shi, W.~Wang, G.-H.~Yang, F.-S.~Yu, and R.~Zhu, Phys. Rev. D \textbf{105}, L031902 (2022), arXiv:2108.06716 [hep-ph].
\bibitem{KiselevEtAl:1999} V.~V.~Kiselev, A.~K.~Likhoded, and A.~I.~Onishchenko, Phys. Rev. D \textbf{60}, 014007 (1999), arXiv:hep-ph/9807354.
\bibitem{KiselevEtAl:2000} V.~V.~Kiselev, A.~K.~Likhoded, and A.~I.~Onishchenko, Eur. Phys. J. C \textbf{16}, 461 (2000), arXiv:hep-ph/9901224.
\bibitem{FaesslerEtAl:2001} A.~Faessler, T.~Gutsche, M.~A.~Ivanov, J.~G.~Korner, and V.~E.~Lyubovitskij, Phys. Lett. B \textbf{518}, 55 (2001), arXiv:hep-ph/0107205.
\bibitem{HernandezEtAl:2008} E.~Hern\'andez, J.~Nieves, and J.~M.~Verde-Velasco, Phys. Lett. B \textbf{663}, 234 (2008), arXiv:0710.1186 [hep-ph].
\bibitem{FaesslerEtAl:2009} A.~Faessler, T.~Gutsche, M.~A.~Ivanov, J.~G.~Korner, and V.~E.~Lyubovitskij, Phys. Rev. D \textbf{80}, 034025 (2009), arXiv:0907.0563 [hep-ph].
\bibitem{AlbertusEtAl:2007} C.~Albertus, E.~Hernandez, J.~Nieves, and J.~M.~Verde-Velasco, Eur. Phys. J. A \textbf{32}, 183 (2007), arXiv:hep-ph/0610030.
\bibitem{BigiUraltsevVainshtein:1992} I.~I.~Bigi, N.~G.~Uraltsev, and A.~I.~Vainshtein, Phys. Lett. B \textbf{293}, 430 (1992), arXiv:hep-ph/9207214.
\bibitem{BigiEtAl:1993} I.~I.~Bigi, M.~A.~Shifman, N.~G.~Uraltsev, and A.~I.~Vainshtein, Phys. Rev. Lett. \textbf{71}, 496 (1993), arXiv:hep-ph/9304225.
\bibitem{NeubertSachrajda:1997} M.~Neubert and C.~T.~Sachrajda, Nucl. Phys. B \textbf{483}, 339 (1997), arXiv:hep-ph/9603202.
\bibitem{GuberinaMelicStefancic:1999} B.~Guberina, B.~Melic, and H.~Stefancic, Eur. Phys. J. C \textbf{9}, 213 (1999), arXiv:hep-ph/9901323.
\bibitem{ChengShi:2018} H.-Y.~Cheng and Y.-J.~Shi, Phys. Rev. D \textbf{98}, 113005 (2018), arXiv:1809.08102 [hep-ph].
\bibitem{LHCb:2023XibcSearch} R.~Aaij \emph{et al.} (LHCb), Chin. Phys. C \textbf{47}, 093001 (2023), arXiv:2204.09541 [hep-ex].
\bibitem{BerezhnoyEtAl:1998} A.~V.~Berezhnoy, V.~V.~Kiselev, A.~K.~Likhoded, and A.~I.~Onishchenko, Phys. Rev. D \textbf{57}, 4385 (1998), arXiv:hep-ph/9710339.
\bibitem{Baranov:1996} S.~P.~Baranov, Phys. Rev. D \textbf{54}, 3228 (1996).
\bibitem{BerezhnoyEtAl:2018} A.~V.~Berezhnoy, A.~K.~Likhoded, and A.~V.~Luchinsky, Phys. Rev. D \textbf{98}, 113004 (2018), arXiv:1809.10058 [hep-ph].
\bibitem{ChangEtAl:2006} C.-H.~Chang, C.-D.~Chen, C.-F.~Qiao, and J.-X.~Wang, Phys. Rev. D \textbf{73}, 094022 (2006), arXiv:hep-ph/0601032.
\bibitem{JiangQiao:2012} X.-G.~Jiang and C.-F.~Qiao, Phys. Rev. D \textbf{86}, 054031 (2012), arXiv:1208.3051 [hep-ph].
\bibitem{ZhangEtAl:2011} J.-W.~Zhang, X.-G.~Wu, T.~Zhong, Y.~Yu, and Z.-Y.~Fang, Phys. Rev. D \textbf{83}, 034026 (2011), arXiv:1101.1130 [hep-ph].
\bibitem{RobertsPervin:2009} W.~Roberts and M.~Pervin, Int. J. Mod. Phys. A \textbf{24}, 2401 (2009), arXiv:0803.3350 [nucl-th].
\bibitem{AlbertusEtAl:2010a} C.~Albertus, E.~Hern\'andez, and J.~Nieves, Phys. Lett. B \textbf{683}, 21 (2010), arXiv:0911.0889 [hep-ph].
\bibitem{AlbertusEtAl:2010b} C.~Albertus, E.~Hernandez, and J.~Nieves, Phys. Lett. B \textbf{690}, 265 (2010), arXiv:1004.3154 [hep-ph].
\bibitem{AlievAziziSavci:2012} T.~M.~Aliev, K.~Azizi, and M.~Savc\i{}, J. Phys. G \textbf{39}, 085006 (2012), arXiv:1205.6320 [hep-ph].
\bibitem{EbertFaustovGalkin:2004} D.~Ebert, R.~N.~Faustov, and V.~O.~Galkin, Phys. Rev. D \textbf{70}, 014018 (2004), arXiv:hep-ph/0404280.
\bibitem{RidgwayWise:2019} A.~K.~Ridgway and M.~B.~Wise, Phys. Lett. B \textbf{793}, 181 (2019), arXiv:1902.04582 [hep-ph].
\bibitem{Yang:2026} G.-H.~Yang, arXiv:2602.00720 [hep-ph] (2026).
\bibitem{EichtenEtAl:1975} E.~Eichten, K.~Gottfried, T.~Kinoshita, J.~Kogut, K.~D.~Lane, and T.-M.~Yan, Phys. Rev. Lett. \textbf{34}, 369 (1975).
\bibitem{EichtenEtAl:1978} E.~Eichten, K.~Gottfried, T.~Kinoshita, K.~D.~Lane, and T.-M.~Yan, Phys. Rev. D \textbf{17}, 3090 (1978).
\bibitem{QuiggRosner:1979} C.~Quigg and J.~L.~Rosner, Phys. Rept. \textbf{56}, 167 (1979).
\bibitem{BuchmullerTye:1981} W.~Buchmuller and S.~H.~H.~Tye, Phys. Rev. D \textbf{24}, 132 (1981).
\bibitem{Martin:1980} A.~Martin, Phys. Lett. B \textbf{93}, 338 (1980).
\bibitem{GodfreyIsgur:1985} S.~Godfrey and N.~Isgur, Phys. Rev. D \textbf{32}, 189 (1985).
\bibitem{LuchaSchoberlGromes:1991} W.~Lucha, F.~F.~Schoberl, and D.~Gromes, Phys. Rept. \textbf{200}, 127 (1991).
\bibitem{WangYuZhao:2017} W.~Wang, F.-S.~Yu, and Z.-X.~Zhao, Eur. Phys. J. C \textbf{77}, 781 (2017), arXiv:1707.02834 [hep-ph].
\bibitem{XiaoEtAl:2017} L.-Y.~Xiao, K.-L.~Wang, Q.-F.~L\"u, X.-H.~Zhong, and S.-L.~Zhu, Phys. Rev. D \textbf{96}, 094005 (2017), arXiv:1708.04384 [hep-ph].
\bibitem{XiaoLuZhu:2018} L.-Y.~Xiao, Q.-F.~L\"u, and S.-L.~Zhu, Phys. Rev. D \textbf{97}, 074005 (2018), arXiv:1712.07295 [hep-ph].
\bibitem{Edmonds:1957} A.~R.~Edmonds, \emph{Angular Momentum in Quantum Mechanics} (Princeton University Press, 1957).
\bibitem{Rose:1957} M.~E.~Rose, \emph{Elementary Theory of Angular Momentum} (Wiley, 1957).
\bibitem{EndoEtAl:2025} M.~Endo, S.~Iguro, S.~Mishima, and R.~Watanabe, ``Constructing heavy-quark sum rule for $b\to c$ meson and baryon decays,'' arXiv:2509.02006 [hep-ph].
\end{thebibliography}
\end{document}